\documentclass[3p,times]{elsarticle}

\usepackage{lineno,hyperref}
\usepackage{floatrow}

\modulolinenumbers[5]

\journal{Annals of Physics}

\biboptions{sort&compress}

\usepackage{amssymb,amsmath,textcomp}
\usepackage{float}
\usepackage{bm}

\usepackage{graphicx}
\usepackage{epsfig,epsf}
\usepackage{epstopdf}

\usepackage{makeidx}

\usepackage{color}
\usepackage{hyperref}
\usepackage{caption}
\usepackage{subcaption}

\usepackage{setspace}

\makeindex

\newcommand{\beq}{\begin{eqnarray}}
	\newcommand{\eeq}{\end{eqnarray}}

\def\be{\begin{equation}}
	\def\ee{\end{equation}}

\newcommand\eq[1]{Eq.~(\ref{#1})}

\def\di{\mathrm{d}}

\numberwithin{equation}{section}

\begin{document}

%
%
%
%
%

\begin{frontmatter}

	\title{\centering{Casimir versus Helmholtz forces: Exact results}}
	\author[DMD,DMD2,JR]{D. M. Dantchev\corref{CorrespondingAuthor}}
	\ead{daniel@imbm.bas.bg}
	\author[NT]{N. S. Tonchev}
	\ead{nicholay.tonchev@gmail.com}
	\author[JR]{J. Rudnick}
	\ead{jrudnickucla@gmail.com}
	
	\address[DMD]{Institute of
		Mechanics, Bulgarian Academy of Sciences, Academic Georgy Bonchev St. building 4,
		1113 Sofia, Bulgaria}
		\address[DMD2]{Max-Planck-Institut f\"{u}r Intelligente Systeme, Heisenbergstrasse 3, D-70569 Stuttgart, Germany}
	\address[NT]{Institute of Solid State Physics, Bulgarian Academy of Sciences,1784 Sofia, Bulgaria} 
	\address[JR]{Department of Physics and Astronomy, University of California, Los Angeles, CA 90095}
	
	\cortext[CorrespondingAuthor]{Corresponding author}

	
	\begin{abstract}
		Recently, attention has turned to the issue of the ensemble dependence of fluctuation induced forces. As a  noteworthy example, in $O(n)$ systems the statistical mechanics underlying such forces can be shown to differ in the constant $\vec{M}$ magnetic canonical ensemble (CE) from those in the widely-studied constant $\vec{h}$ grand canonical ensemble (GCE). Here, the counterpart of the Casimir force in the GCE is the \textit{Helmholtz} force in the CE.  Given the difference between the two ensembles  for finite systems, it is reasonable to anticipate that these forces will have, in general, different behavior for the same geometry and boundary conditions. Here we present some exact results for both the Casimir and the Helmholtz force in the case of the one-dimensional Ising model subject to periodic and antiperiodic boundary conditions and compare their behavior. We note that the Ising model has  recently being solved in Phys.Rev. E {\bf 106} L042103
		(2022), using a combinatorial approach, for the case of fixed value $M$ of its order parameter.  Here we derive exact result for the partition function of the one-dimensional Ising model of $N$ spins and fixed value $M$ using the transfer matrix method (TMM); earlier results obtained via the TMM were limited to $M=0$ and $N$ even. As a byproduct, we derive several specific  integral representations of the hypergeometric function of Gauss. Using those results, we rigorously derive that the free energies of the CE and grand GCE are related to each other via Legendre transformation in the thermodynamic limit, and  establish the leading finite-size corrections for the canonical case, which turn out to be much more pronounced than the corresponding ones in the case of the GCE.
	\end{abstract}
	
	\date{\today}
	
\end{frontmatter}




\section{Introduction}

In any ensemble --- grand canonical, canonical, or micro-canonical --- one can define a thermodynamic fluctuation induced force which is specific for that ensemble. Since these ensembles correspond to quite different physical conditions, it is reasonable to
expect that the behavior of these forces differ between ensembles. However, the precise
behavior of these forces has not yet been the object of thorough and systematic
study.

In this article we discuss the Casimir force, pertinent to the grand canonical
ensemble (GCE), and the Helmholtz force, pertinent to the canonical ensemble (CE), and will
compare their behavior as a function of the temperature.   We will do that for periodic (PBC) and antiperiodic (ABC) boundary conditions. Explicitly, we will demonstrate that for the Ising chain with length $L$, fixed value of the total magnetization $M$ under PBC the Helmholtz force $F_{\rm H}^{\rm (per)}(L,T,M)$ can, depending on the temperature $T$, be attractive or repulsive, while the Casimir force $F_{\rm Cas}^{\rm (per)}(L,T,h)$, with $h$ being external magnetic field, is only attractive.  This issue is by no means limited to the Ising chain and can be addressed, in principle, in any model of interest. The analysis reported here can also be viewed as a useful addition to approaches to fluctuation-induced forces in the fixed-$M$ ensemble based on Ginzburg-Landau-Wilson Hamiltonians \cite{RSVG2019,GVGD2016,GGD2017} in which the authors studied the "canonical Casimir force", which, in fact, coincides with the Helmholtz force defined below.

In many cases, actual calculations utilize the ensemble  which appears to be the most convenient, often  regardless  of that how it reflects the physical situation.   This is based on the principle of equivalence of assemblies. The equivalence of ensembles follows from  the key procedure of thermodynamic limit  --- i.e., for infinite (bulk) systems ensembles see, e.g., \cite{Ru99,Minlos2000}.   However, this equivalence is not applicable in the case of fluctuation
induced forces which result from the imposed boundary conditions, i.e., for finite systems. In such a case, one has no reason to expect the same behavior of the fluctuation-induced forces in different ensembles. This was announced in a recent Letter \cite{DR2022}, where an analogue of the Casimir force,  named  the Helmholtz force, was attributed to the canonical ensemble. 

Let us envisage a system with a film geometry $\infty^{d-1}\times L$, $L\equiv L_\perp$, and with boundary conditions $\zeta$ imposed along the spatial direction of finite extent $L$.  Take ${\cal F}_{ {\rm tot}}^{(\zeta)}$ to be the total free energy of such a system within the GCE. Then, if   $f^{(\zeta)}(T,h,L)\equiv \lim_{A\to\infty}{\cal F}_{ {\rm tot}}^{(\zeta)}/A$  is the free energy per area $A$ of the system, one can define the Casimir force for critical systems in the grand-canonical $(T-h)$-ensemble, see, e.g. Ref. \cite{Dantchev2023}, as: 
\begin{equation}
	\label{CasDef}
	\beta F_{\rm Cas}^{(\zeta)}(L,T,h)\equiv- \frac{\partial}{\partial L}f_{\rm ex}^{(\zeta)}(L,T,h)
\end{equation}
where
\begin{equation}
	\label{excess_free_energy_definition}
	f_{\rm ex}^{(\zeta)}(L,T,h) \equiv f^{(\zeta)}(L,T,h)-L f_b(T,h)
\end{equation}
is the so-called excess (over the bulk) free energy per area and per $\beta^{-1}=k_B T$.

Along these lines we define the corresponding Helmholtz  fluctuation induced force in canonical $(T-M)$-ensemble:
\begin{equation}
	\label{HelmDef}
	\beta F_{\rm H}^{(\zeta)}(L,T,M)\equiv- \frac{\partial}{\partial L}f_{\rm ex}^{(\zeta)}(L,T,M)
\end{equation}
where
\begin{equation}
	\label{excess_free_energy_definition_M}
	f_{\rm ex}^{(\zeta)}(L,T,M) \equiv f^{(\zeta)}(L,T,M)-L f_H(T,m),
\end{equation}
with $m=\lim_{L, A\to \infty}M/(LA)$, and $f_H$ is the Helmholtz free energy density of the ``bulk'' system. In the remainder we will take $L=N a$, where $N$ is an integer number, and for simplicity we 
set $a=1$, i.e., all lengths will be measured in units of the lattice spacing $a$. 

We will show that the Helmholtz fluctuation induced force  has a behavior very different from that of the Casimir force.

In Ref. \cite{DR2022} the long standing problem for the exact solution of the one--dimensional Ising chain with fixed value of the order parameter $M$ has been solved. The approach undertaken there was a combinatorial one. It has been successfully applied for the case of PBC, ABC and Dirichlet-Dirichlet boundary conditions.
For the PBC case of a chain with $N$ spins interacting with coupling $K$ the following result was derived 
\begin{equation}
	\label{eq:Z-final-introduction-per}
	Z^{(\rm per)}_C(N,K,M) =	N e^{K (N-4)} \, _2F_1\left(\frac{1}{2} (-M-N+2),\frac{1}{2}
	(M-N+2);2;e^{-4 K}\right),\quad |M|<N,
\end{equation}
where $_2F_1(\alpha,\beta;\gamma;z)$ is the generalized Gauss  hypergeometric function \cite{AS}. For ABC the corresponding result reads 
\begin{eqnarray}
		\label{eq:Z-final-introduction-antiper}
	Z^{(\rm anti)}_C(N,K,M)
	& = &e^{K (N-6)} \left[2 \left(e^{4 K}-1\right) \, _2F_1\left(\frac{1}{2} (-M-N+2),\frac{1}{2}
	(M-N+2);1;e^{-4 K}\right) \right. \nonumber \\ & & \left. +N \, _2F_1\left(\frac{1}{2} (-M-N+2),\frac{1}{2}
	(M-N+2);2;e^{-4 K}\right)\right]. 
\end{eqnarray}
On the other hand, attempts to apply one of the best known approaches usually used for solving both the one and two-dimensional Ising model in GCE, namely the transfer matrix method, has led, despite some attempts, to limited success. Indeed, in the fixed $M$ CE it has proven successful, see Ref.  \cite{Henkel2008}, just for the case $M=0$. For the partition function of a system of $2N$ particles with coupling $K$ the authors of \cite[p.305, Eq. S4]{Henkel2008}  obtained 
\begin{equation}
	\label{eq:Henkel-M=0}
		Z^{(\rm per)}_C(2N,K,M=0)=(2\sinh(2K))^N[P_N(\coth 2K)-P_{N-1}(\coth 2K)], 
\end{equation}
where $P_N(x)$ are the Legendre polynomials. 
For the case of arbitrary $M$ the authors derived
\cite[p.307, Eq. S5]{Henkel2008} 
\begin{equation}
	\label{eq:Henkel-general-M}
	Z^{\rm (per)}_C(N,K,M)=2Ne^{2KN}\sum_{l=0}^{N}\frac{1}{2N-l}\left( \begin{array}{c} 2N-l \\ N+M/2\end{array} \right) \left( \begin{array}{c} N+M/2 \\ l \end{array} \right)\left(\frac{1-e^{4K}}{e^{4K}}\right)^l. 
\end{equation}

In this article we demonstrate how to solve the one-dimensional Ising model with fixed $M$ via the transfer matrix method. We also show how to derive 	\eq{eq:Henkel-M=0} starting from \eq{eq:Z-final-introduction-per}, and demonstrate how to obtain \eq{eq:Z-final-introduction-per} starting from \eq{eq:Henkel-general-M}. Based on the transfer matrix approach, we have also derived an integral form representation of 	$Z^{(\rm per)}_C(N,K,M)$ and 	$Z^{(\rm anti)}_C(N,K,M)$, which allows us to obtain the leading behavior of the partition function and related Helmholtz free energy. This, along with the corresponding results for the Ising chain within the GCE,  allows us to verify the validity of the Legendre transformation  connecting the Helmholtz and Gibbs free energies, as well as to elucidate the type of leading $N$ corrections to the corresponding bulk results. We have established that while in the GCE the corrections to the bulk behavior for $N\gg 1$ are, as expected, exponentially small, for the CE they are of the order of $\ln (N)/N$. In deriving this, we have obtained, as a byproduct, a few new integral representations of specific hypergeometric functions.

Let us stress, as noted in Refs. \cite{DR2022,Henkel2008}, that in
applications of the equilibrium Ising
model to binary alloys or binary liquids, which one customarily considers, if one insists on full
rigor, the case with $M$ fixed must be addressed. In such systems the meaning of $M$ is the dominance of one of the components over the other one. We note that recently one dimensional and quasi one-dimensional systems have been the objects of intensified experimental interest---see, e.g., \cite{Balandin2022} and references cited therein. Some of these, like ${\rm TaSe_3}$, are
quasi-one dimensional in the sense that they have strong covalent bonds in one
direction along the atomic chains and weaker bonds in the perpendicular
plane \cite{Stolyarov2016}. Others are
more properly considered true one dimensional materials, in that they have covalent bonds only
along the atomic chains and much weaker van der Waals interactions
in perpendicular directions \cite{Balandin2022}. One-dimensional van der Waals materials have emerged as an entirely new field, which encompasses interdisciplinary work by physicists, chemists, materials
scientists, and engineers \cite{Balandin2022}. The Ising chain considered here can be seen as the simplest possible example of such a one-dimensional material. Feynman checkers \cite{gersch1981feynman}, also known as a one-dimensional quantum walk (or Hadamard walk), can
non-trivially be viewed as one-dimensional Ising model with imaginary temperature and fixed ”magnetization”, e.g., 
in the canonical ensemble. The current literature on the subject is quite extensive, see,  e.g. Ref. \cite{Skopenkov2020}, 
and Refs. therein. 

As wee see from the above, despite the fact that the one-dimensional Ising model is a topic that has been around for a very long time, publication activity on it in different contexts is still very active - see, e.g., \cite{gersch1981feynman,hanneken1998exact,Antal2004,denisov2005domain,Nandhini2009,GarciaPelayo2009,kofinger2010single,taherkhani2011investigation,XC2019,sitarachu2020exact,Skopenkov2020,maniwa2003one}.

The structure of the current article is as follows. In Sec. \ref{sec:periodic} we consider PBC. We present the results for the  Gibbs free  energy in Sec. \ref{sec:grand-canonical-ensemble} and in Sec. \ref{sec:periodic-canonical}  for the Helmholtz free energy.  We express our findings in terms of Chebyshev polynomials of first and second kind. The equivalence of GCE and CE and their leading $N$ dependencies is discussed in Sect. \ref{sec:canonical-versus-grand-canonical}. In Sec. \ref{sec:antiperiodic-bc} the Helmholtz and Gibbs free energies for the finite chain under ABC are derived and discussed. Then 
Sec. \ref{sec:ensemble-dependent-fluctuation-induced-forces} is devoted to the behavior Casimir and Helmholtz fluctuation induced forces under PBC and ABC.  The mathematical techniques needed to follow the derivation of the results reported in the article are given in the appendices. \ref{sec:M=0 case} shows how to derive from our results \eq{eq:Henkel-M=0}, while \ref{sec:Henkel-general-M}  demonstrates how from 	\eq{eq:Henkel-general-M} to obtain	\eq{eq:Z-final-introduction-per}. \ref{appendix:asymptotic} contains the derivation of the leading finite–size behavior of the Helmholtz free energy. The mathematical details needed for Sec. \ref{sec:antiperiodic-bc} and Sec. \ref{sec:ensemble-dependent-fluctuation-induced-forces} are given in \ref{section:integral-hypergeometric} and \ref{eq:sec:general_limiting_form}. In \ref{section:integral-hypergeometric} an algebra of the Gauss hypergeometric functions helpful for the study
of the integral representations of the PBC and ABC canonical statistical sums has been developed. In \ref{eq:sec:general_limiting_form} the question of the scaling behavior of the Helmholtz force and its leading corrections have been clarified. The article closes with concluding remarks and discussion given in Sec. \ref{sec:summary}.   
	
\section{The Helmholtz and Gibbs free energies for the finite chain  under PBC} 
\label{sec:periodic}

In this section, for PBC, we consider the behavior of the Gibbs free energy in the GCE and the Helmholtz free energy in the CE. For PBC the results for the Gibbs free energy can be found in many books of statistical mechanics --- see, e.g., \cite{B82,H87,PB2011,Berlinsky2019}. We report them here in order to introduce the notations to be used throughout the current article and to show that they can be also  expressed in terms of Chebyshev polynomials of the first kind. 

\subsection{The case of the grand canonical ensemble}
\label{sec:grand-canonical-ensemble}

Let $N$ is the total number of spins in the one dimensional nearest neighbor interaction  Ising chain with PBC, $S_{N+1}=S_1$, $S=+1$ or $S=-1$  and let $h\in \mathbb{R}$ be the dimensionless magnetic field. 

For the free energy density of the grand canonical ensemble one has
\beq
\label{eq:free_energy_per}
\beta f^{\rm (per)}(N,K,h)
= -\frac{1}{N}{Z}^{\rm (per)}_{\rm GC}(N,K,h) 
=-\ln[\lambda_1(K,h)]-\frac{1}{N}\ln\left\{1+\exp[-N/\xi(K,h)]\right\},
\eeq
where 
\begin{equation}
	\label{npr}
	{Z}^{\rm (per)}_{\rm GC}(N,K,h)=
	\sum_{\left\{S_i\right\}}\exp\bigg[K\sum_{(i,j)}S_iS_j+h\sum_iS_i \bigg]
	=\lambda_1^N(K,h)+\lambda_2^N(K,h)	
\end{equation}	
is the the grand canonical statistical sum, and 
\be
\label{eq:eigenvalues_per1}
\lambda_{1,2}(K, h) = e^K \left(\cosh h \pm \sqrt{e^{-4 K} +  \sinh^2 h}\right). 
\ee
In \eq{eq:free_energy_per} we have used that, see, e.g., Ref. \cite[p. 36]{B82}
\be
\label{eq:scaling_length_Ising}
\xi^{-1}(K,h)=\ln \left[\lambda_1(K,h)/\lambda_2(K,h)\right].
\ee
Since $\lambda_1(K, h)=\lambda_1(K, -h)$ and $\lambda_2(K, h)=\lambda_2(K, -h)$, one has $f^{\rm (per)}(N,K,h)=f^{\rm (per)}(N,K,-h)$. Due to this fact in all what follows we will assume, for simplicity, that $h\ge 0$.

From \eq{eq:free_energy_per} one immediately obtains 
\beq
\label{eq:free_energy_per-bul-and-excess}
\beta f^{\rm (per)}_b(K,h) =-\ln[\lambda_1(K,h)] \quad \mbox{and} \quad N \beta [f^{\rm (per)}(N,K,h)- f^{\rm (per)}_b(K,h)]= -\ln\left\{1+\exp[-N/\xi(K,h)]\right\}.
\eeq

From here on it will prove useful to have another representation of $\beta f^{\rm (per)}(N,K,h)$ in terms of Chebyshev  polynomials $T_n(z)$ \cite[Eq. (1.49)]{mason2002chebyshev} of the first kind 
\begin{equation}
	\label{cp11}
	T_N(z)=\frac{1}{2}\left[(z+\sqrt{z^2-1})^N + (z-\sqrt{z^2-1})^N\right], \quad z\in {\mathbb C}.
\end{equation}
One has 
\begin{equation}
	\label{eq:free_energy_pern1}
	{Z}^{\rm (per)}_{\rm GC}(N,K,h)
	=2\left(\sqrt{2\sinh(2K)}\right)^N T_N\left(x(K,h\right)) \quad \mbox{and} \quad 
	\beta f^{\rm (per)}(N,K,h)
	= -\frac{1}{N}\ln \left\{2\, r^N(K)T_N\left[x(K,h)\right]\right\},
\end{equation}
where the shorthand notation $ r(K):=\sqrt{2\sinh(2K)}$ is used
and
\begin{equation}
	\label{Nsv}
	x(K,h)=\frac{e^K\cosh(h)}{\sqrt{2\sinh(2K)}}>1,\quad\forall h\in [0,\infty),K>0.
\end{equation}

\subsection{The case of the canonical ensemble}
\label{sec:periodic-canonical}

Let us denote by $M$ the total magnetization of the system and  again let $N$ be the total number of spins (particles) in the Ising chain. We are interested in the Helmholtz free energy density 
\begin{equation}
	\label{eq:Helmh-def}
	\beta a^{(\rm per)}(N,K,m)=-\ln{Z^{(\rm per)}_C}(N,K,m)/N, \quad \mbox{where}\quad N>0,\; N\in \mathbb{N} \quad \mbox{and} \quad M\in\mathbb{N},\; M\in[-N,N],
\end{equation}
of the system, where the statistical sum in the canonical ensemble  is given by
\begin{equation}
	Z^{(\rm per)}_C(N,K,m)=\sum_{\left\{S_i\right\}} e^{K{\cal H}}\delta_{S,Nm}, \quad \mbox{where} \quad 
	\quad S\equiv \sum_{i=1}^N S_i, \quad {\cal H}=\sum_{(i,j)}S_iS_j,
\end{equation}
and we have introduced the magnetization per spin
(particle) $m=M/N \in [-1,1]$. Since $S_i=\pm 1$, it is easy to show that $N$ and $M$ have the \textit{same} parity. 

Using the integral representation of the Kronecker delta function 
\begin{equation}
	\delta_{S,N m}=\frac{1}{2\pi}\int_{-\pi}^{+\pi}e^{i(S-Nm)\phi}d\phi
\end{equation}
($ "S"$ and $"N\times m=M" $ are integers) one obtains 
\begin{eqnarray}
	\label{eq:transfer-matrix-periodic}
	Z^{(\rm per)}_C(N,K,m)=\frac{1}{2\pi}\int_{-\pi}^{\pi}\sum_{\{S_i\}} e^{-K {\cal H}+i\phi\left(\sum_{n} S_n-Nm\right)}d\phi =\frac{1}{2\pi}\int_{-\pi}^{\pi}{e^{-i N m \phi}\left[\sum_{\{S_n\}} e^{-K {\cal H}+i\phi\sum_{n} S_n}\right]} d\phi.
\end{eqnarray}
Note that the expression in the rectangular brackets, Eq.\eqref{eq:transfer-matrix-periodic}, is identical to the grand canonical statistical sum  in a complex field ${\tilde h}=i\phi$. Thus the following sequence of equations holds:
\begin{align}
	\label{eq:complex-lambda}
	\lambda_{1,2} (K,i\phi)\equiv \hat{\lambda}_{1,2}(K,\phi)  =e^K\left[\cos{\phi}\pm i\sqrt{{\sin}^2{\phi}-e^{-4K}}\right],
\end{align}
and
\begin{equation}
	\label{eq:sum-of-lambdas}
	{Z}^{\rm (per)}_{\rm GC}(N,K,i\phi)=\hat{\lambda}_1^N(K,\phi)+\hat{\lambda}_2^N(K,\phi).
\end{equation}
As a result, from \eq{eq:free_energy_pern1} we get
\begin{equation}
	\label{lambdasa}
	{Z}^{\rm (per)}_{\rm GC}(N,K,i\phi)
	=2\left(\sqrt{2\sinh(2K)}\right)^N T_N\left(x(K,i\phi\right)),
\end{equation}
where (cf. Eq.\eqref{Nsv})
\begin{equation}
	\label{Nsh}
	x(K,i\phi)\equiv\tilde {x}(K,\phi):=\frac{e^K\cos(\phi)}{\sqrt{2\sinh(2K)}}=
	\frac{\cos(\phi)}{\sqrt{1-e^{-4K}}}, \quad K>0.
\end{equation}
Inserting Eq.\eqref{lambdasa} in Eq.\eqref{eq:transfer-matrix-periodic}
we obtain 
\begin{eqnarray}
	\label{eq:transfer-matrix-periodics}
	Z^{(\rm per)}_C(N,K,m)=\frac{1}{\pi}\left(\sqrt{2\sinh(2K)}\right)^N\int_{-\pi}^{\pi}{e^{-i N m\phi}\left[ T_N\left(x(K,i\phi\right))\right]} d\phi.
\end{eqnarray}
Since $\tilde {x}(K,-\phi)=\tilde {x}(K,\phi)$, we have
\begin{eqnarray}
	\label{eq:transfer-matrix-periodicse}
	Z^{(\rm per)}_C(N,K,m)=\frac{2}{\pi}\left(\sqrt{2\sinh(2K)}\right)^N\int_{0}^{\pi}\cos (N m \phi) T_N\left(\tilde{x}(K,\phi)\right)) d\phi,
\end{eqnarray}
Because of the symmetry property $T_N(-x)=T_N(x)$ for $N$ even, and $T_N(-x)=-T_N(x)$ for $N$ odd, and because of the equality of the parity of $N$ and $M=N m$ it is easy to show that the canonical statistical sum is:
\begin{equation}
	\label{eq:Z-fixed-m-via-angle-pi-over-two2}
	Z^{(\rm per)}_C(N,K,m) =\frac{4}{\pi} \; r^N(K)\int _{0}^{\pi/2}{\cos(N m \phi)}   T_N\bigg(\frac{  \cos (\phi)}{\sqrt{1-e^{-4 K}}}\bigg)\; \di \phi, \quad \mbox{where} \quad r(K)= \sqrt{2 \sinh (2 K)}>0.
\end{equation}
From Eq.\eqref{eq:Z-fixed-m-via-angle-pi-over-two2}, for the Helmholtz free energy density $a^{(\rm per)}(N,K,m)$ we obtain 
\begin{align}
	\label{eq:H-free-energy}
	\beta a^{(\rm per)}(N,K,m)=-\ln \left[r(K)\right]-\frac{1}{N}\ln\left\{{\cal I}(N,K,m)\right\},
\end{align}
where
\begin{equation}
	\label{Tch1}
	{\cal I}(N,K,m)=\frac{4}{\pi} \; \int _{0}^{\pi/2}{\cos(N m \phi)}   T_N\bigg(\frac{  \cos (\phi)}{\sqrt{1-e^{-4 K}}}\bigg)\; \di \phi=
	\frac{4}{\pi} \; \int _{0}^{\pi/2}{\cos(N m x)   \cos\left[N \varphi (K,x)\right]\; \di x}.
\end{equation}
Here the trigonometric presentation of the Chebyshev polynomials $T_N(x) $
\begin{equation}
	\label{D34}
	T_N(x)_= \left\{\begin{array}{cc}\cos(N\cos^{-1}(x)), & |x|  \leq 1\\  
		\\\cosh(N\cosh^{-1}(x)), & \qquad x\geq 1, \qquad \end{array}\right., 
\end{equation}
is used, and  the function 
\begin{equation}
	\label{eq:varphi-function}
	\varphi(K,x)=\arccos\left(\frac{\cos(x)}{\sqrt{1-e^{-4K}}}\right)
\end{equation}
has been introduced. The above expressions can be easily evaluated, of course, numerically.

In \ref{section:integral-hypergeometric} we prove that  	\eq{eq:Z-fixed-m-via-angle-pi-over-two2} and \eq{eq:Z-final-introduction-per} are two different representations of the same quantity. 
%
%
The preference for one or the other is a matter of convenience. For example 
in the derivation of the leading size behavior of the Helmholtz free energy  the integral representation of the statistical sum will be used, see  \ref{appendix:asymptotic}. The representation based on the Gauss hypergeometric function is convenient, on its turn,  for the analysis of the high-temperature and low-temperature behavior of the ensemble dependent fluctuation induced  Helmholtz force, as well as for it finite-size scaling behavior.

\subsection{On the equivalence of the ensembles and on their large $N$ behavior}
\label{sec:canonical-versus-grand-canonical}

Let us clarify the question about the equivalence of the grand canonical and canonical ensembles for the finite Ising chain. By doing so, we will also determine the large $N$ corrections to the corresponding Gibbs and Helmholtz free energies of the finite chain. 

We start by the obvious relation
\begin{equation}
	\label{eq:relation-grand-canonical}
	Z^{\rm (per)}_{\rm GC}(N,K,h)=\int_{-N}^{N} dM \exp\left[h\, M\right]\; Z^{\rm (per)}_C(N,K,M).
\end{equation}
Using it and applying the steepest descent  method (see Ref. \cite{F77,Fedoryuk1987}), in \ref{appendix:asymptotic} it is shown that the asymptotic behavior of ${\cal I}(N,K,m)$ (see 	\eq{Tch1}) for $N\gg 1$ is
\begin{eqnarray}\label{eq:asymptote}
	{\cal I}(N,K,m)\simeq\frac{4  }{\sqrt{2 \pi  N \chi_b}} \; \exp  \left\{N\left[-  h m+  \ln \left(\frac{\sqrt{1-\left(1-e^{-4 K}\right) m^2}+e^{-2 K}}{\sqrt{\left(1-e^{-4 K}\right) \left(1-m^2\right)}}\right) \right]\right\}\left[1+O(N^{-1})\right].
\end{eqnarray}
For the bulk magnetization, using \eq{eq:free_energy_per-bul-and-excess}, one has 
\begin{equation}\label{eq:bulk-magnetization}
	m_b(K,h):=-\partial[\beta f_b(K,h)]/\partial h
=\frac{1}{\sqrt{1+\left[e^{2 K}\sinh(h)\right]^{-2}}}.
\end{equation}
\eq{eq:bulk-magnetization} can be inverted to obtain 
\be
\label{eq:hb-versus-m}
h_b(K,m_b)=\sinh^{-1}\left(\frac{e^{-2 K} m_b}{\sqrt{1-m_b^2}}\right),
\ee
and, thus, 
\begin{equation}\label{chi-b}
	\chi_b\equiv \chi_b(K,m_b)\equiv \partial m_b(K,h_b)/\partial h_b=e^{2 K} \left(1-m_b^2\right) \sqrt{1-\left(1-e^{-4 K}\right) m_b^2}.
\end{equation}
Using the above results, one can easily determine  the corresponding thermodynamic potential $a_b(K,m_b)$.
According to Legendre transformation 
\be
\label{eq:helmhotz_free_energy_bulk}
\beta a_b(K,m_b)=\beta f_b(K,h(K,m_b))+h(K,m_b) m_b.
\ee
Thus, the bulk Helmholtz free energy can be obtained in explicit form
\begin{align}
	\label{eq:Helmholtz_free_energy_bul_explicit}
	\beta a_b(K,m_b) = K + \frac{1}{2}\ln(1 - m_b^2) - \ln\left[1 + \sqrt{m_b^2 + e^{4 K} (1 - m_b^2)}\right]  + \; m_b \sinh ^{-1}\left(\frac{e^{-2 K} m_b}{\sqrt{1-m_b^2}}\right).
\end{align}
For the scaling behavior of $\beta a_b(K,m_b)$, i.e., for $K\gg 1$, from 	\eq{eq:Helmholtz_free_energy_bul_explicit} we obtain 
\begin{equation}
\label{eq:ab-scaling}
\beta a_b(K,m_b)\simeq -K -\exp[-2K]\sqrt{1-m_b^2}, \quad K \gg 1. 
\end{equation}
 
 In accordance with principle of equivalence of ensembles \eq{eq:Helmholtz_free_energy_bul_explicit} must be in agreement with
the result that follows from the one of the Helmholtz free energy of the finite system after taking the thermodynamic
limit. In \ref{appendix:asymptotic} we prove that this is correct and determine the leading $N$-dependent corrections to the bulk result. From Eqs.  \eqref{eq:H-free-energy}, \eqref{eq:asymptote}, and  \eqref{eq:Helmholtz_free_energy_bul_explicit} it follows that the leading $N$-dependent correction to the bulk Helmholtz free energy is (see also Ref. \cite{Henkel2008})
\begin{equation}
\label{eq:corrections-Helmholtz}
\beta a^{(\rm per)}(N,K,m)-	\beta a_b(K,m)= {\cal O}(\ln N/N),
\end{equation}
while from Eqs. \eqref{eq:free_energy_per} and \eqref{eq:free_energy_per-bul-and-excess} it follows that  the leading $N$-dependent correction to the bulk Gibbs free energy is
\begin{equation}
\label{eq:corrections-Gibbs}
\beta f^{(\rm per)}(N,K,h)-	\beta f_b(K,h)= {\cal O}\left(\exp(-N/\xi(K,h))\right), \quad \mbox{when}
\quad \xi(K,h)={\cal O}(1).
\end{equation}
The last implies that, for PBC, finite-size corrections are much more profound in the CE than in the GCE where, as expected, they are  exponentially small.

\section{The Helmholtz and Gibbs free energies for the finite chain under ABC} 
\label{sec:antiperiodic-bc}

In this section, following the recipe applied above, we consider the behavior of the Gibbs free energy in the GCE and the Helmholtz free energy in the CE for ABC. The results for the Helmholtz free energy by means of a combinatorial procedure can be found in Ref. \cite{DR2022}. Here, by
means of transfer matrix method and using the language of Chebyshev polynomials, we derive the partition function of the Ising chain  for both GCE and CE. 

\subsection{The case of the grand canonical ensemble}
\label{sec:grand-canonical}

In the case of ABC, as distinct from the PBC case, at some position of the chain the coupling instead of being $K$ is changed to $-K$. The transfer matrix which reflects such a coupling we denote by ${ B}_{A}$. Thus the partition function of the one-dimensional Ising model in a magnetic field is given by 
\begin{equation}
	{Z}^{\rm (anti)}_{\rm GC}(N,K,h) 
	= \mathop{\rm Tr} ( { B}^{N-1} \cdot { B}_{A} )
=\mathop{\rm Tr} \left[\left({P}^{-1}\cdot{ {B}}^{N-1} \cdot{P}\right)\cdot \left({P}^{-1} \cdot {  B}_{A}\cdot  {P}\right)\right]. \label{eq:JR1}
\end{equation}
Explicitly one has
\begin{equation}
	\label{eq:Z-A-partition-function}
	{Z}^{\rm (anti)}_{\rm GC}(N,K,h) = \mathop{\rm Tr} 
	\left[
	\begin{pmatrix}
		\lambda_{1}^{N-1}(K,h) & 0 \\ \\
		0 & \lambda_{2}^{N-1}(K,h)  
	\end{pmatrix} \cdot {P}^{-1} \cdot \begin{pmatrix}
		\exp\left(K+h\right) & \exp\left(-K\right) \\ \\
		\exp\left(-K\right) & \exp\left(K-h \right)    
	\end{pmatrix}\cdot  {P}
	\right].
\end{equation}
Where we have taken into account that the transfer matrix ${B}_A$ appropriate to an antiferromagnetic bond is
\begin{equation}
	{B}_A = \left(  \begin{array}{cc} e^{-K+ h} & e^K \\ e^K &e^{-K- h}  \end{array}\right). \label{eq:JR2}
\end{equation}
The matrix $P$, which diagonalizes ${B}$, can be written in the form 
\begin{equation}
	\label{eq:orthogonal_matrix_P}
	{P}=\begin{pmatrix}
		\cos\phi & -\sin \phi \\ \\
		\sin \phi & \cos \phi  
	\end{pmatrix}, \quad \mbox{with} \quad {P}^{-1}\hat{B}{P}=\begin{pmatrix}
		\lambda_{1}(K,h) & 0 \\ \\
		0 & \lambda_2(K,h) 
	\end{pmatrix},
\end{equation}
where $\phi$ is such that, see Ref. \cite{B82}
\begin{equation}
	\label{eq:def-phi}
	\cot 2\phi=\exp(2K)\sinh h, \quad 0<\phi<\pi/2.
\end{equation}
Explicitly, one has
\begin{eqnarray}
	\label{eq:matrix-P-explicit}
	\cos \phi= \frac{1}{\sqrt{2}}\sqrt{1+\frac{\sinh (h)}{\sqrt{\sinh ^2(h)+e^{-4 K}}}} \quad \mbox{and} \quad 
	\sin \phi= \frac{1}{\sqrt{2}}\sqrt{1-\frac{\sinh (h)}{\sqrt{\sinh ^2(h)+e^{-4 K}}}}.
\end{eqnarray}
Performing the calculations in \eq{eq:Z-A-partition-function}, we obtain 
\begin{equation}
	\label{eq:partition-anti-final}
	{Z}^{\rm (anti)}_{\rm GC}(N,K,h) = \frac{\cosh (h) }{\sqrt{e^{4 K} \sinh ^2(h)+1}}\left[\lambda _1^N(K,h)-\lambda _2^N(K,h)\right].
\end{equation}

From \eq{eq:scaling_length_Ising} and \eq{eq:partition-anti-final} one obtains
\begin{eqnarray}
	\label{eq:scaling-aper}
	{Z}^{\rm (anti)}_{\rm GC}(N,K,h) =  \frac{\cosh (h) }{\sqrt{e^{4 K} \sinh ^2(h)+1}}\lambda_1^N(K,h) \left(1-e^{-N/\xi(K,h)}\right). 
\end{eqnarray}
From Eqs.  \eqref{eq:free_energy_per-bul-and-excess} and \eqref{eq:scaling-aper} it follows that 
\begin{equation}
	\label{eq:corrections-Gibbs-bulk-finite}
	N[\beta f^{(\rm anti)}(N,K,h)-	\beta f_b(K,h)]= -\ln \left[\cosh (h) \bigg/\sqrt{e^{4 K} \sinh ^2(h)+1}\right] -\ln\left\{1-\exp[-N/\xi(K,h)]\right\}.
\end{equation}
The last implies that the finite-size corrections are, as expected, dominated by a term of the order of ${\cal O}(1)$ that represents the interface free energy of the interface effected under such boundary conditions,  plus exponentially small (if $\xi(K,h)={\cal O}(1)$) corrections in grand canonical ensembles.

It is useful to have another representation of $\beta f^{\rm (anti)}(N,K,h)$ in terms of Chebyshev  polynomials of the second kind $U_n(z)$ \cite[Eq. (1.52)]{mason2002chebyshev}
	\begin{equation}
		\label{cp116}
		U_{N-1}(z)=\frac{1}{2\sqrt{z^2-1}}\left[(z+\sqrt{z^2-1})^N -          (z-\sqrt{z^2-1})^N\right], \quad z\in {\mathbb C}.
	\end{equation}
	Using \eq{eq:partition-anti-final}, \eq{eq:eigenvalues_per1}, and \eq{Nsv} one has 
	\begin{eqnarray}
		\label{eq:free_e_antiU}
		{Z}^{\rm (anti)}_{\rm GC}(N,K,h)
		=2e^{-2K}\left(\sqrt{2\sinh(2K)}\right)^N x(K,h)U_{N-1}\left(x(K,h\right)),
	\end{eqnarray}
	and
	\begin{eqnarray}
		\label{eq:free_energy_antiU}
		\beta f^{\rm (anti)}(N,K,h)
		= -\frac{1}{N}\ln \left\{2\, r^N(K)U_{N-1}\left[x(K,h)\right]\right\}
		-\frac{1}{N}\ln[e^{2K}x(K,h)].
	\end{eqnarray}
	Note that \eq{eq:free_energy_antiU} for $\beta f^{\rm (anti)}(N,K,h)$ is identical in form (up to corrections of order $O(1/N)$,) to \eq{eq:free_energy_pern1} for  $\beta f^{\rm (per)}(N,K,h)$. The different boundary conditions (ABC and
	PBC) yield the different polynomials (of the second, or first kind, correspondingly); in the limit $N \to \infty$ the both coincide.

From the many well known mutual recurrence equations for the Chebyshev polynomials, e.g., \cite[Eq. (1.14)]{mason2002chebyshev}
\begin{equation}
	T_{N}(x)=T_{2N}(u),\quad U_{N}(x)=\frac{1}{2}u^{-1}U_{2N+1}(u),\quad u=[(1/2)(1+x)]^{1/2},
\end{equation}
\cite[Eq. (1.3a)]{mason2002chebyshev}
\begin{equation}
	T_{N+1}(x)\,=x\,T_{N}(x)-(1-x^2)\,U_{N-1}(x), \quad T_N(x)\,=\,2x T_{N-1}(x)
		-T_{N-2}(x),
\end{equation}
and \cite[Eq. (1.7)]{mason2002chebyshev} 
\begin{equation}
U_{N+1}(x)\,=\,x U_N(x)  - T_{N+1}(x),\quad U_N(x)\,=\,2x U_{N-1}(x)
		-U_{N-2}(x)
\end{equation}
immediately follows different mutual recurrence relations for the partition functions, e.g., like this one
\begin{eqnarray}
	\label{eq:energy_antiU}
		e^{-2K} x(K,h)Z^{\rm (anti)}_{\rm GC}(N,K,h)
		=\frac{x(K,h)}{1-x(K,h)^2}Z^{\rm (per)}_{\rm GC}(N,K,h) - \left(\sqrt{2\sinh(2K)}\right)\frac{1}{1 -x(K,h)^2 }Z^{\rm (per)}_{\rm GC}(N+1,K,h).
\end{eqnarray}

\subsection{The case of the canonical ensemble}
\label{sec:antiperiodic-canonical}

Here we discuss how, using the results from the previous subsection, to obtain the partition function of the Ising chain in CE within the framework of the transfer matrix
approach.

\subsubsection{Approach based on the transfer matrix calculations}

Similar to what has been done in  Sec. \ref{sec:periodic-canonical} for periodic boundary conditions 
\begin{eqnarray}
	\label{eq:transfer-matrix-antiperiodic}
	Z^{(\rm anti)}_C(N,K,m) &=&\frac{1}{2\pi}\int_{-\pi}^{\pi}{e^{-i N m \phi}\left[\sum_{\{S_i\}^{\rm (anti)}} e^{-\beta H+i \phi \sum_{i} S_i}\right]} d\phi=\frac{1}{2\pi}\int_{-\pi}^{\pi}{e^{-i N m \phi}} \;Z^{(\rm anti)}_{\rm GC}(N,K,i \phi)\,d\phi \\
	&=&\frac{1}{2\pi}\int_{-\pi}^{\pi}{e^{-i N m \phi}} \;\frac{\cos (\phi)}{\sqrt{1-e^{4 K} \sin ^2(\phi)}}\left[\hat{\lambda} _1^N(K,\phi)-\hat{\lambda}_2^N (K,\phi)\right]\,d\phi, \nonumber
\end{eqnarray}
where the set of spins $\{S_i\}^{\rm (anti)}$ obey ABC, and $N m=M\in \mathbb{Z}$ is an integer number, and we used that the integrand is an even function of $x$. 

Using \eq{eq:free_e_antiU}, \eq{eq:complex-lambda}, and  \eq{Nsh}, we can rewrite \eq{eq:transfer-matrix-antiperiodic}  in the form
	\begin{equation}
	\label{ICh2}
		Z^{(\rm anti)}_C(N,K,m)=\frac{4}{\pi}e^{-2K}\left(\sqrt{2\sinh(2K)}
		\right)^N\int_{0}^{\pi/2}{e^{-i N m \phi}} \ x(K,i\phi)U_{N-1}\left(x(K,i\phi\right))\,d\phi,
	\end{equation}
	where we took into account that  $N$ and $M$ are of the same parity, and thus the integrand is symmetric with respect to $\pi/2$.
	Now, we will use the trigonometric representation of $U_N(z)$
	\begin{equation}
		\label{D35}
		U_{N-1}(x)_= \left\{\begin{array}{cc}\sin(N\cos^{-1}(x))/
			\sin(\cos^{-1}(x)), & |x|  \leq 1\\  
			\\\sinh(N\cosh^{-1}(x))/\sinh(\cosh^{-1}(x)), & \qquad x\geq 1, \qquad \end{array}\right., 
	\end{equation}
	and \eq{eq:varphi-function}, which leads to  
	\begin{eqnarray}
		\label{eq:transfer-matrix-antiperiodic-final-simple}
		Z^{(\rm anti)}_C(N,K,M) =  \left[ r(K)
		\right]^Ne^{-2K}	{\rm  D}(N,K,M),  \quad \mbox{where} \quad r(K)= \sqrt{2 \sinh (2 K)}>0.
	\end{eqnarray}
Here we have introduced the notation
	\begin{equation}
	{\rm  D}(N,K,M):=\frac{4}{\pi}\int_0^{\pi/2}\frac{\cos(Mx)
			\cos(x)}{\sqrt{ \sin ^2(x)-e^{-4 K}}}
		\sin(N\varphi(K,x) )\; dx.
	\end{equation}
	From Eq.\eqref{eq:transfer-matrix-antiperiodic-final-simple}, for the Helmholtz free energy density $a^{(\rm anti)}(N,K,m)$ we obtain 
	\begin{align}
\label{eq:H-free-energy-D}
\beta a^{(\rm anti)}(N,K,m)=-\ln( r(K))-\frac{1}{N}\ln\left\{{\rm  D}(N,K,mN)\right\} +\frac{2K}{N},
\end{align}

In \ref{section:integral-hypergeometric} we prove that  	\eqref{ICh2} and Eq.\eqref{eq:Z-final-introduction-antiper} are two different representations of the same quantity. Similar to the situation with the periodic boundary conditions case, the preference for one or the other is a matter of convenience.
Also in \ref{section:integral-hypergeometric},  we prove an  another representation of the statistical sum:  
\begin{equation}
	\label{zzantiper}
	Z^{\rm (anti)}_C(N,K,M)={\cal A}(N,K,M)\,Z^{\rm (per)}_C(N,K,M)\,+\,{\cal B}(N,K,M)\,{Z^{\rm (per)}_C}(N-2,K,M),
\end{equation}
where
\begin{equation}
	\label{DAB}
	{\cal A}(N,K,M) := \left[e^{-2K}+\frac{N^2-M^2}{N(N-1)}\sinh(2K)\right],
	\qquad {\cal B}(N,K,M):=-2\frac{(N-2)^2-M^2 }{(N-1) (N-2)} \sinh
	^2(2 K).
\end{equation}
Thus, the statistical sum for ABC can be expressed in terms of the ones for PBC. 

\subsubsection{Approach based on the combinatorial calculations}
\label{sec:antiperiodic-canonical-combinatorial-approach}

Let us note that this result may be obtained  from the study of the ABC statistical sum in \cite{DR2022}
where  it is shown analyzing the microscopic states of the system that $Z^{{\rm (anti)}}_C(N,K,M)$ may be rewritten , almost obviously, in the form:
\begin{equation}
	\label{eq:DR_zanti}
	Z^{{\rm (anti)}}_C(N,K,M)=e^{-2K}Z^{\rm (per)}_C(N,K,M)+4\sinh(2K)e^{K(N-4)}\\_2F_1\left(\frac{1}{2}(-M-N+2),\frac{1}{2}(M-N+2);1,e^{-4K}\right).
\end{equation}
After that, using the identity \cite[ Eq. (2.25), with $c=2$]{Rakha2011}
we obtain \eq{zzantiper}.

\section{Ensemble dependent fluctuation induced forces}
\label{sec:ensemble-dependent-fluctuation-induced-forces}

In any ensemble one can define a fluctuation induced force which is specific for that ensemble. How the behavior of these forces differ from each other is a problems that has not yet been  studied in depth or detail. In the current article we will study the Casimir force, pertinent to grand canonical ensemble, and the Helmholtz force, pertinent to the canonical ensemble, and will compare their behavior as a function of the temperature.   We will do that for periodic and antiperiodic boundary conditions. 
Explicitly, we will demonstrate that for the Ising chain with fixed $M$ under PBC $F_{\rm H}^{\rm (per)}(T,M,L)$ can, depending on the temperature $T$, be attractive or repulsive, while $ F_{\rm Cas}^{\rm (per)}(T,h,L)$ is only attractive.  This issue is by no means limited to the Ising chain and can be addressed, in principle, in any model of interest. The analysis reported here can also be viewed as a useful addition to approaches to fluctuation-induced forces in the fixed-$M$ ensemble based on Ginzburg-Landau-Wilson Hamiltonians \cite{RSVG2019,GVGD2016,GGD2017} in which one studied the usual Casimir force.

\subsection{Casimir force}

We start with the case of PBC. 

\subsubsection{The case of PBC}
\label{sec:ensemble-dependent-fluctuation-induced-forces-periodic}

From \eq{eq:free_energy_per}, and  \eq{eq:free_energy_per-bul-and-excess}, for the Casimir force in the ensemble with fixed external boundary field $h$ we derive
\beq
\label{eq:Cas_periodic_Ising}
\beta F_{\rm Cas}^{(\rm per)}(N,K,h)
= -\frac{1}{N}\frac{N}{\xi (K,h)}\frac{e^{-N/\xi (K,h)}}{ 1+e^{-N/\xi (K,h)}}.
\eeq

Note that \eq{eq:Cas_periodic_Ising} is consistent with scaling behavior of the Casimir force under PBC for \textit{any} $K$ and $h$ in terms of $N/\xi(K,h)$. Furthermore, it follows that $F_{\rm Cas}^{(\rm per)}(N,K,h)<0$, i.e., the force is \textit{attractive}, again for \textit{any} value of $N,K,h$. 

%

Starting from \eq{eq:scaling_length_Ising}, one can easily identify the scaling variables. First, it is clear that $\xi$ diverges when $\lambda_2\to\lambda_{1}$. Obviously, this happens when $h\to 0$ and $K\to\infty$. Defining
\be
\label{eq:xit_Ising}
\xi_t\equiv\xi(K,0)\simeq \frac{1}{2}e^{2K}, \;\mbox{when}\; K\gg 1, \mbox{and}\; \xi_h\equiv\lim_{K\to \infty}\xi(K,h)\simeq \frac{1}{2h}, \; \mbox{when} \; h\ll 1,
\ee
for the scaling variables one identifies
\be\label{eq:scaling_variables_Ising}
x_t=N/\xi_t= 2 N e^{-2K}, \quad \mbox{and} \quad x_h =N/\xi_h=2 N h. 
\ee
Thus, in terms of these scaling variables the correlation length, the bulk magnetization, and the bulk Gibbs free energy in the limit $K\gg 1$ read 
\be
\label{eq:corr_length_scaling}
\xi(K,h)=\frac{N}{\sqrt{x_h^2+x_t^2}}, \quad m_b(K,h)=\frac{x_h}{\sqrt{x_h^2+x_t^2}}, \quad \beta f_b(K,h)= -K-\frac{1}{4 N}\sqrt{x_h^2+x_t^2}.
\ee
Recalling that in terms of $T_c=0$,  $t=\exp(-2K)$ and $h$, one can define the usual scaling relations with \cite{B82}
\be
\label{eq:crit_exponents}
\alpha=\gamma=\nu=\eta=1,\; \beta=0,\;\delta=\infty, \; \mbox{but so that}\; \beta \delta=1. 
\ee
For completeness, we note that in the limit $K\to 0$ one has
\be
\label{eq:Kto0}
\xi(0,h)=0,\quad m_b(K,h)=\tanh(h),\quad f_b(0,h)=-\ln[2 \cosh (h)].
\ee
Note that these limiting values can not be achieved as a limit of the corresponding scaling functions given in \eq{eq:corr_length_scaling}.

\begin{figure}[h!]
	\centering
	\includegraphics[width=3.0in]{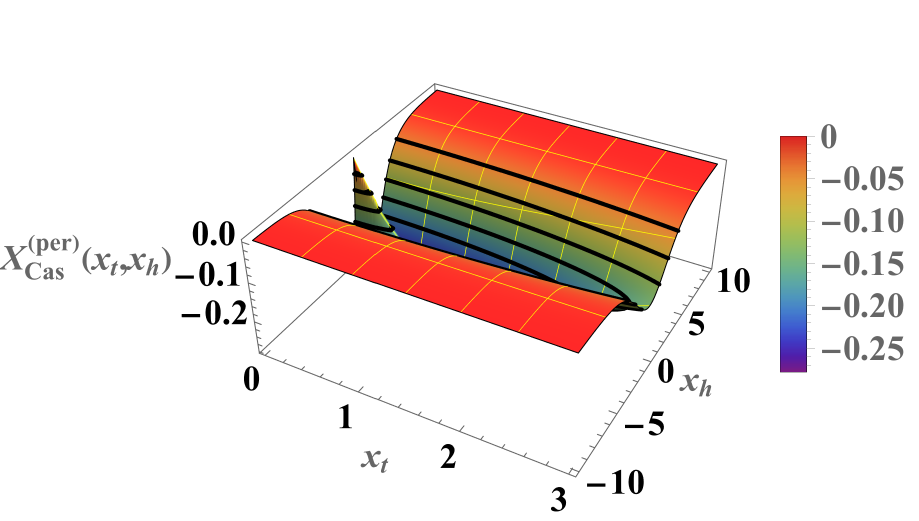} \quad
	\includegraphics[width=3.0in]{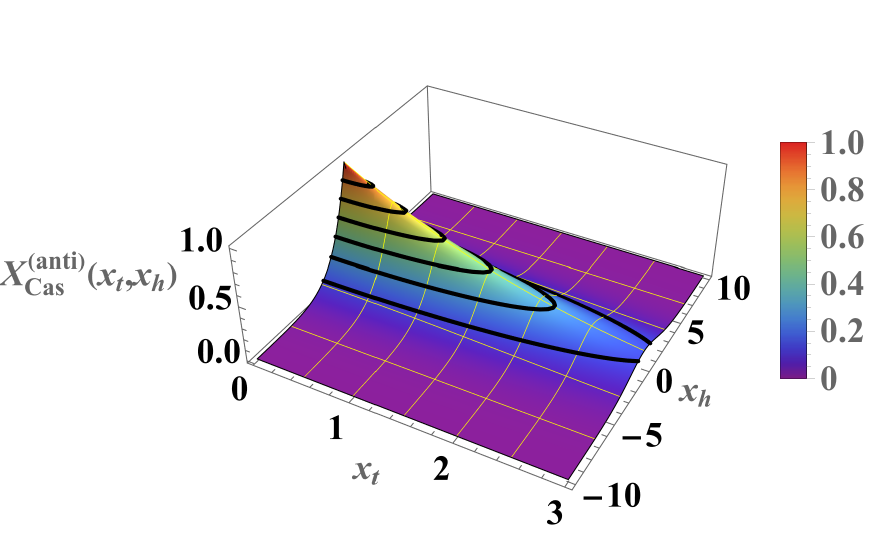}
	\caption{The behavior of the scaling function $X_{\rm Cas}^{(\rm per)}(x_t,x_h)$.   Relief plots of the scaling function of the Casimir force as a function of $x_t$ and $x_h$. The left panel shows the behavior of function $X_{\rm Cas}^{\rm (per)}(x_t,x_h)$ for PBC (see 	\eq{eq:fCas_Isng_scalig}). We observe that the function is always \textit{negative}, corresponding to \textit{attractive} fore, and symmetric about $h=0$. The right panel shows the behavior  of  $X_{\rm Cas}^{\rm (anti)}(x_t,m)$ (see \eq{eq:fCas_Isng_scalig-antiperiodic-t-h}) for ABC. The force, contrary to the periodic case, in always \textit{repulsive}. }
	\label{fig:XCas_per_anti_xt_xh_3D}
\end{figure}

Now, from \eq{eq:Cas_periodic_Ising} and using \eq{eq:corr_length_scaling} for $\beta F_{\rm Cas}^{(\rm per)}(N,K,h)$ one obtains the Casimir force in terms of the usual scaling variables 
\beq
\label{eq:fCas_Isng_scalig}
\beta F_{\rm Cas}^{(\rm per)}(N,K,h)&=&\frac{1}{N}X_{\rm Cas}^{(\rm per)}(x_t,x_h), \quad \mbox{where} \quad 
X_{\rm Cas}^{(\rm per)}(x_t,x_h)\; =\; -\sqrt{x_h^2+x_t^2}\; \frac{\exp\left[-\sqrt{x_h^2+x_t^2}\right]}{1+\exp\left[-\sqrt{x_h^2+x_t^2}\right]}.
\eeq
At the essential critical point $(T=0,h=0)$ one has $\lim_{x_t\to 0, x_h\to 0} X_{\rm Cas}^{(\rm per)}(x_t,x_h)=0$, which is the maximal value of the force. The behavior of $X_{\rm Cas}^{(\rm per)}(x_t,x_h)$ is visualized in Fig. \ref{fig:XCas_per_anti_xt_xh_3D}.

\subsection{The case of ABC }

From	\eq{eq:corrections-Gibbs-bulk-finite} for the Casimir force in the ensemble with fixed external boundary field $h$ under ABC  we derive
\beq
\label{eq:Cas_antiperiodic_Ising}
\beta F_{\rm Cas}^{(\rm anti)}(N,K,h)
= \frac{1}{N} \frac{N}{\xi (K,h)} \frac{e^{-N/\xi (K,h)}}{ 1-e^{-N/\xi (K,h)}}>0.
\eeq
Obviously
\beq
\label{eq:fCas_Isng_scalig-antiperiodic-t-h}
\beta F_{\rm Cas}^{(\rm anti)}(N,K,h)&=&\frac{1}{N}X_{\rm Cas}^{(\rm anti)}(x_t,x_h), \quad \mbox{where} \quad 
X_{\rm Cas}^{(\rm anti)}(x_t,x_h) =\sqrt{x_h^2+x_t^2} \frac{\exp\left[-\sqrt{x_h^2+x_t^2}\right]}{1-\exp\left[\sqrt{x_h^2+x_t^2}\right]}>0,
\eeq
i.e., this force is always \textit{repulsive}. At the essential critical point one has $\lim_{x_t\to 0, x_h\to 0} X_{\rm Cas}^{(\rm anti)}(x_t,x_h)=1$, which is the maximal value of the force. The behavior of the force is shown in Fig. \ref{fig:XCas_per_anti_xt_xh_3D}. 

Note that there is a simple relation between the Casimir force under periodic and antiperiodic boundary conditions --- see \eq{eq:Cas_periodic_Ising} and \eq{eq:Cas_antiperiodic_Ising}:
\begin{equation}
	\label{eq:relation-periodic-antiperiodic}
	\beta F_{\rm Cas}^{(\rm per)}(N,K,h) = \beta F_{\rm Cas}^{(\rm anti)}(2N,K,h)-\beta F_{\rm Cas}^{(\rm anti)}(N,K,h) .
\end{equation}

Now we continue with the study of the Helmholtz force. We will observe that in this case there is no such a simple relation as given in \eq{eq:relation-periodic-antiperiodic}.

\subsection{Helmholtz force}

In the introduction we defined the \textit{Helmholtz} fluctuation induced force - see Eqs.  \eqref{HelmDef} and \eqref{excess_free_energy_definition_M}. 
We will show that the so-defined  force  has a behavior very different from that of the Casimir force.

\subsubsection{The case of PBC}

The statistical sum of the model is presented in  	\eq{eq:Z-final-introduction-per}, 
provided $0\le M<N$. This expression can be also written in the form of polynomial, since both $(-N-M+2)/2$ and $(M-N+2)/2$ are negative integers. The expression for $_2F_1$ can be written in the form
\begin{equation}
	\label{eq:Z-in-terms-of-Np-Nm}
	_2F_1\left(1-N_+,1-N_-;2;e^{-4 K}\right)=\sum _ {k = 0}^{N_- - 1}\binom {N_+ - 1} {k}\binom {N_- - 
		1} {k}\frac {e^{(-4 K) k}} {k + 1},
\end{equation}
where $N_+=(N+M)/2$ is the number of positive spins in the system, while  $N_-=(N-M)/2$ is the number of negative spins. In the scaling regime $K\gg 1, \; N\gg 1$ with $x_t=2 N e^{-2K}={\cal O}(1)$, see  \eq{eq:scaling_variables_Ising}, one can obtain a representation of the $_2F_1$ function in a scaling form. From \eq{eq:Z-in-terms-of-Np-Nm} one has 
\begin{eqnarray}
		\label{eq:Z-in-terms-of-Np-Nm-to-scaling-second}
&&_2F_1\left(-N_++1,
		-N_- +1;2;e^{-4 K}\right)=\sum_{k=0}^{N_-}(N_+-1)(N_--1)\cdots (N_+-k)(N_--k)\frac{e^{-(4K)k}}{(k+1)(k!)^2}\nonumber \\
	&\simeq&\sum_{k=0}^{\infty}\frac{(1-m^2)^k x_t^{2k}}{4^{2k}(k+1)(k!)^2}=\frac{4 I_1\left[\sqrt{1-m^2}\;x_t/2\right]}{ \sqrt{1-m^2} \;x_t},
\end{eqnarray}
which leads to the replacement of the hypergeometric function $_2F_1$ with the scaling function $	F_s(m,x_t)$, i.e., 
\begin{equation}
	\label{eq:F-scaling}
	_2F_1 \to F_s(m,x_t)=\frac{4 I_1\left[\sqrt{1-m^2}\;x_t/2\right]}{ \sqrt{1-m^2} \;x_t}=\frac{2 I_1\left[\sqrt{1-m^2} N \exp (-2 K)\right]}{ \sqrt{1-m^2} N \exp (-2 K)}.
\end{equation}
Here $I_1(z)$ is  modified Bessel function of the first kind. In the asymptotic regimes $x_t\to 0$, and $x_t\to \infty$, one derives 
\begin{equation}
	\label{eq:F-asymptotes}
	F_s(m,x_t)\simeq\left\{\begin{array}{lll}
		1+\frac{1}{32} \left(1-m^2\right) x_t^2, & x_t\to 0\\
		& \\
		\frac{4 }{\sqrt{\pi } \left(\sqrt{1-m^2} x_t\right)^{3/2}}\exp{\left[\sqrt{1-m^2} x_t/2\right]}, & x_t\to \infty,
	\end{array}\right.
\end{equation}
respectively. When deriving \eq{eq:Z-in-terms-of-Np-Nm-to-scaling-second} from 	\eq{eq:Z-in-terms-of-Np-Nm} we have neglected the numbers of the order $(-k+1)$ in the multipliers.  This, for small $k$, can be safely done in all terms in which $N_-$ is not very small. For large $k$ the approximation can be safely utilized again, because then the factor $\exp[-4Kk]/(k+1)(k!)^2$ strongly suppresses the contribution of such terms, especially for $K\gg 1$.

From 	\eq{eq:Z-final-introduction-per} for the Helmholtz free energy one arrives at the exact expression
\begin{eqnarray}
	\label{eq:Helmholtz-free-energy}
	\beta a^{(\rm per)}(N,K,M) = -\frac{1}{N}\ln \left[\, _2F_1\left(\frac{1}{2} (-M-N+2),\frac{1}{2} (M-N+2);2;e^{-4 K}\right)\right] -K+\frac{4 K}{N}-\frac{\ln (N)}{N}.
\end{eqnarray}

From \eq{eq:F-asymptotes} one concludes that the behavior of $\beta a^{(\rm per)}(N,K,m)$ for $x_t\gg 1$ is consistent with that one of $\beta a_b(K,m_b)$ for $K\gg 1$, see \eq{eq:Helmholtz_free_energy_bul_explicit}.
For the scaling behavior of the $a^{(\rm per)}(N,K,m)$ one obtains 
\begin{eqnarray}
	\label{eq:Hfe_scaling}
	\beta a^{(\rm per)}_s(N,K,m)= -K-\frac{1}{N}\left[\ln (N)-4 K+\ln\frac{4 I_1\left(\sqrt{1-m^2} x_t/2\right)}{x_t \sqrt{1-m^2}}\right]+{\cal O}(N^{-2}).
\end{eqnarray}
Obviously, the scaling is not perfect, i.e., it is violated, because of the existence of a logarithmic in $N$ term. This terms exist always when $M<N$.

From \eq{eq:Hfe_scaling} one can derive an analytical expression for the scaling function $X_H(x_t,m)$
of the Helmholtz force
\begin{equation}
	\label{eq:fH-scaling}
	F_{H}^{(\rm per)}(N,K,m)\simeq \frac{1}{N} X_H^{\rm (per)}(x_t,m),
\end{equation}
where
\begin{equation}
	\label{eq:XH-scaling}
	X_H^{\rm (per)}(x_t,m)=-\frac{1}{2} \sqrt{1-m^2} x_t+\frac{x_t I_0\left(\frac{1}{2} x_t \sqrt{1-m^2}\right)}{2 \sqrt{1-m^2} I_1\left(\frac{1}{2} x_t \sqrt{1-m^2}\right)}-\frac{1+m^2}{1-m^2}.
\end{equation}
In the limit $K\gg 1$, and $N$ such that $x_t\ll 1$, one obtains
\be
\label{eq:H-force-scaling}
X_{H}^{(\rm per)}(x_t,m) \simeq \left\{\begin{array}{cc}0, & |m|=1\\
1-\frac{1}{2} \sqrt{1-m^2} x_t+\frac{1}{8} \left(1-m^2\right) x_t^2, &|m|<1.\end{array}\right. 
\ee
Realizing that
\be
1-\frac{1}{2} \sqrt{1-m^2} \;x_t+\frac{1}{8} \left(1-m^2\right) x_t^2=\left(1-\frac{1}{4} \sqrt{1-m^2}\; x_t\right)^2+\frac{1}{16} \left(1-m^2\right) x_t^2,
\ee
we conclude that the scaling function is \textit{positive} for small values of $x_t$ for any $|m|<1$. When $x_t\gg 1$ one has
\begin{equation}
	\label{eq:as-H-per}
X_{H}^{(\rm per)}(x_t,m)\simeq x_t/2+1/(2 \sqrt{1-m^2}).
\end{equation}

\begin{figure}[h!]
	\centering
	\includegraphics[width=3.0in]{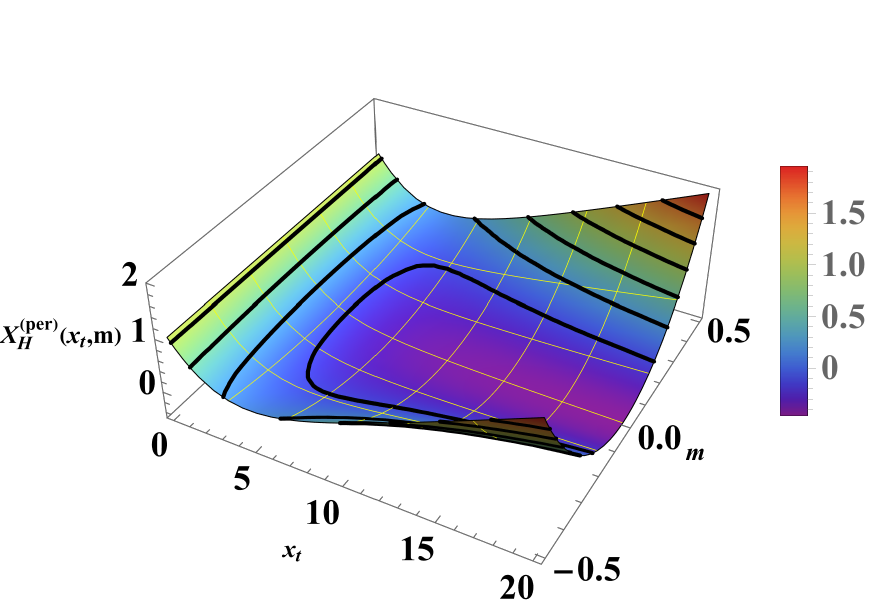} \quad
	\includegraphics[width=3.0in]{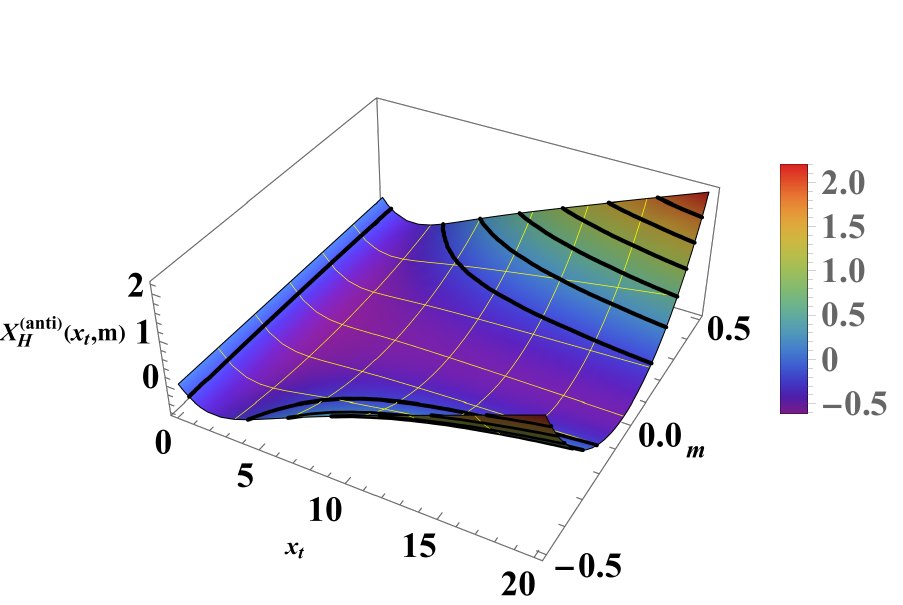}
	\caption{Relief plots of the scaling function of the Helmholtz force as a function of $x_t$ and $m$ for PBC. The left panel shows the behavior of function $X_H^{\rm (per)}(x_t,m)$ for PBC (see 	\eq{eq:XH-scaling}), while the right panel shows the behavior  of  $X_H^{\rm (anti)}(x_t,m)$ for ABC.}
	\label{fig:Helmholtz3D}
\end{figure}

\subsubsection{The case of antiperiodic boundary conditions }
 The statistical sum in this case is given in \eq{eq:Z-final-introduction-antiper}. 
Performing the expansion for large values of $K$ in this expression after expressing $M$ as $m N$, and $K$ as  $K=\ln[2N/x_t]/2$ (see \eq{eq:scaling_variables_Ising}), one obtains \eq	{eq:Z-in-terms-of-Np-Nm-to-scaling-second} and
\begin{equation}
	\label{eq:scaling-forms-second}
	\, _2F_1\left(\frac{1}{2} (-M-N+2),\frac{1}{2} (M-N+2);1;e^{-4 K}\right)\to I_0\left(\frac{1}{2} \sqrt{1-m^2} x_t\right),
\end{equation}
see \eq{eq:sec:general_limiting_form} for the mathematical details. 
Setting these results in \eq{eq:Z-final-introduction-antiper} and rearranging the terms in the  resulting expression, one ends up with the result that 
\begin{equation}
	\label{eq:aperiodic-scaling}
	Z^{\rm (anti)}_C=\frac{1}{N} e^{K N} x_t  \left[I_0\left(\frac{1}{2} \sqrt{1-m^2} x_t\right)+\frac{1}{2N}\frac{x_t I_1\left(\frac{1}{2} \sqrt{1-m^2} x_t\right)}{\sqrt{1-m^2}}\right]\left(1+{\cal O}(N^{-1})\right), \quad K\gg 1,\quad N\gg 1, \quad x_t={\cal O}(x_t).
\end{equation}
Thus, for the Helmholtz free energy of the system one arrives at 
\begin{equation}
	\label{eq:helmholtz-aper-bc}
	\beta a^{(a)}[N, K, m] = 
	-K+\frac{\ln (N)}{N}-\frac{1}{N}\ln \left[x_t \; I_0\left(\frac{1}{2} \sqrt{1-m^2} x_t\right)\right]+{\cal O}(N^{-2}).
\end{equation}
We observe, as in the case of periodic boundary conditions, the presence of logarithmic corrections in the finite-size behavior of the free energy, i.e., lack of a perfect scaling. This logarithmic terms is due to the degeneracy in the position of the seam of the antiferromagnetic coupling. 

Given (\ref{HelmDef}), (\ref{excess_free_energy_definition_M}),  and 	\eqref{eq:Helmholtz_free_energy_bul_explicit}, along with (\ref{eq:Z-final-introduction-antiper}) for the case $\zeta$=``anti'' we are in a position to calculate the Helmholtz force in the fixed-$M$ magnetic canonical ensemble. 

It is reasonable to expect that the Helmholtz force will depend on the total magnetization, $M$. In fact the dependence is on the ratio $m=M/N$.  

\begin{figure}[h!]
	\centering
	\includegraphics[width=3.0in]{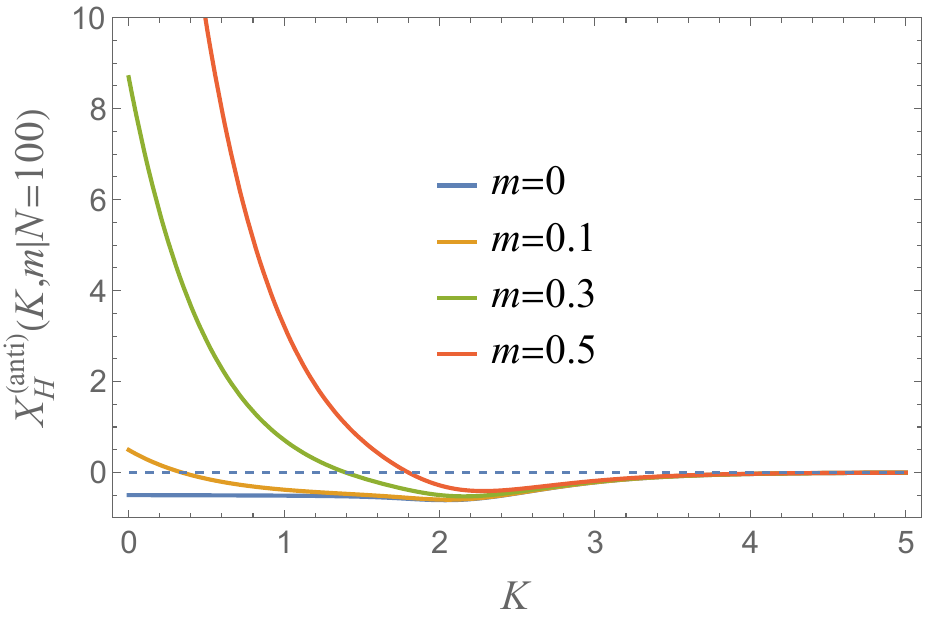} \quad
	\includegraphics[width=3.0in]{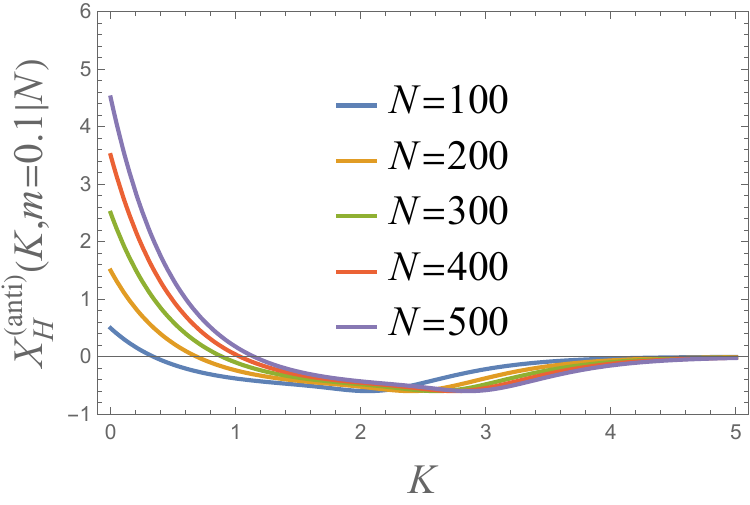}
	\caption{On the left panel: The behavior of the  function $X_{\rm H}^{(\rm anti)}(K,m|N)$  with $N=100$ and for $m=0.1, 0.3$, and $m=0.5$. We observe that the function is negative  for large values of $K$. If $m$ lies below a threshold value depending on $N$ the function is negative for all values of $K$, and if $m$ is greater than that threshold value the function is positive when $K$ is small. On the right panel: The behavior of the function $X_{\rm H}^{(\rm per)}(K,m|N)$ with $N=100,200, 300, 400$ and $N=500$.   Because the value of $m$, set equal to 0.1 here, lies above threshold (see the portion of this caption that refers
		to the left panel) for all values of $N$,  in this plot, the Helmholtz force is attractive for larger $K$ and repulsive for smaller $K$.}
	\label{fig:Helmholtz}
\end{figure}

%

\begin{figure}[htbp]
	\includegraphics[width=\columnwidth]{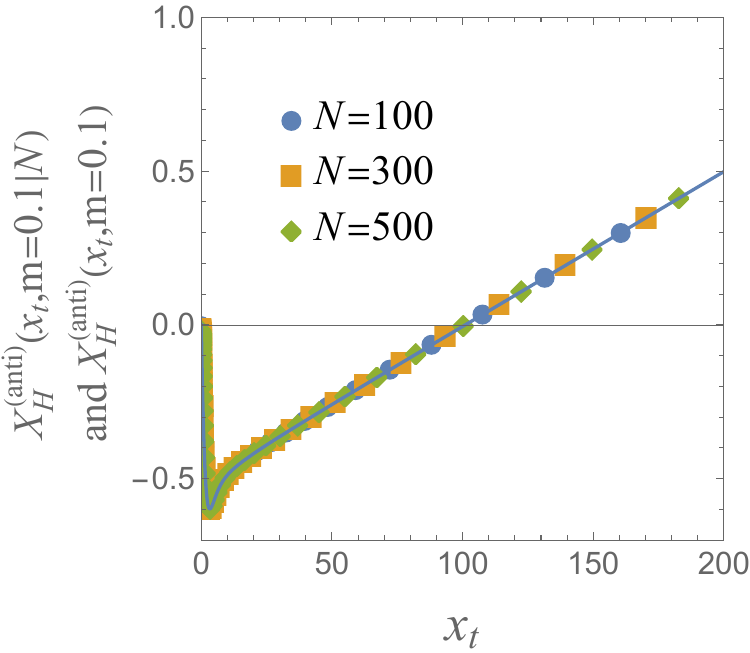}
	\caption{The behavior of the scaling function $X_{\rm H}^{(\rm anti)}(x_t,m|N)=N \times \beta F_{\rm H}^{(\rm anti)}(N,K,m)$ for $m=0.1$ and  $N$=100, 300 and $N=500$, along with the explicit limiting curve $X_{\rm H}^{(\rm anti)}(x_t,m)$ obtained from  \eq{eq:antiscaling2}.  The plot illustrates the scaling behavior of the Helmholtz force when $N$ is large and $x_t=N/\xi_t=2Ne^{-2K}$ (see Eq. (\ref{eq:scaling_variables_Ising})) is not too large. }
	\label{fig:Helmholtz3}
\end{figure}

In the limit of large $N$ and $\xi(K)$ with $x_t=N/\xi(K)$ neither too large nor too small, i.e. the scaling regime, the form of the partition function simplifies and is given in \eq{eq:aperiodic-scaling}. 
Inserting this version of the partition function into (\ref{HelmDef}) and (\ref{excess_free_energy_definition_M}) we end up with the following scaling expression derived from the Helmholtz force
\begin{eqnarray}
	F_H^{(\rm anti)}(K,m)=\frac{1}{N} X_H^{(\rm anti)}(x_t,m),
	\label{eq:antiscaling1}
\end{eqnarray}
where $x_t$ is defined in (\ref{eq:scaling_variables_Ising}). The explicit form of the scaling function $X_H^{(\rm anti)}(x_t,m)$ is
\begin{equation}
	X_{\rm H}^{(\rm anti)}(x_t,m) = \frac{1}{2}\frac{x_t I_1\left(\sqrt{1-m^2} x_t\right)}{\sqrt{1-m^2} I_0\left(\sqrt{1-m^2}
		x_t\right)}-\frac{1}{2}\sqrt{1-m^2} x_t.  
		\label{eq:antiscaling2}
\end{equation}
This expression yields a curve that lies on top of the plots of $N \times \beta F_{\rm H}^{(\rm anti)}(N,K,m)\equiv X_{\rm H}^{(\rm per)}(K,m|N)$ for $N$ equal to 100, 300 and 500 in Fig. \ref{fig:Helmholtz}. 
Finite size  corrections to the partition function and the Helmholtz force go as $1/L$; see 	\eq{eq:helmholtz-aper-bc}. It is easy to show that when $x_t\ll 1$
\begin{equation}
\label{eq:as-XH}
	X_{\rm H}^{(\rm anti)}(x_t,m) \simeq -\frac{1}{2} \sqrt{1-m^2} x_t+\frac{x_t^2}{8},
\end{equation}
while for $x_t\gg 1$
\begin{equation}
	\label{eq:as-XH-large}
	X_{\rm H}^{(\rm anti)}(x_t,m) \simeq \frac{m^2 }{2 \sqrt{1-m^2}}x_t-\frac{1}{2 \left(1-m^2\right)}. 
\end{equation}

Thus, $	X_{\rm H}^{(\rm anti)}(x_t=0,m)=0 $ - see Fig. \ref{fig:Helmholtz3}. For small values of $x_t$ the scaling function $X_{\rm H}^{(\rm anti)}(x_t,m)$ then gets negative (attractive force), then further increases and for larger values of $x_t$ becomes positive (repulsive force). Finally,  for even larger values of $x_t$ it increases linearly - see \eq{eq:as-XH-large}. 

\section{Concluding remarks and discussion}
\label{sec:summary}

In the current article we studied the issue of the ensemble dependence of the fluctuation induced forces. We did that for the Casimir force --- pertinent to the GCE --- when the number of constituents in the system which become critical can vary due to exchange with some reservoir,  and to the Helmholtz force --- when there is a fixed amount of such constituents in the system. The specific calculations have been done for the one dimensional Ising chain under periodic and antiperiodic boundary conditions. The Ising chain in CE has only recently being solved via  a combinatorial approach in Ref. \cite{DR2022}. Here we have extended this study by using the standard transfer matrix approach
in conjunction with the properties of the Chebyshev polynomials. We have demonstrated the versatility of these polynomials in the
description of the finite-size behavior of the model.  Using our approach, we were able to show how some results available in the literature with the transfer method for $M=0$, or intermediate results for $N$ even can be easily derived through our approach, or can be manipulated to the form of the explicit results derived in  Ref. \cite{DR2022}. Among the other results reported in the article are 
\begin{itemize}
	\item We have derived the scaling function of the Casimir force as a function of both the temperature and the magnetic field --- both for PBC and ABC. Somewhat surprisingly this has not been published in the available literature. The results are given by \eq{eq:Cas_periodic_Ising} and \eq{eq:fCas_Isng_scalig}, for PBC, and by \eq{eq:Cas_antiperiodic_Ising} and \eq{eq:fCas_Isng_scalig-antiperiodic-t-h}, for ABC; they are visualized in  Fig. \ref{fig:XCas_per_anti_xt_xh_3D}. As expected, the Casimir force is attractive under PBC and repulsive under ABC. Let us note that, unexpectedly, for the both boundary conditions the scaling functions of the force can be expressed in terms of a single scaling variable --- $N/\xi(K,h)$, where $\xi(T,h)$ is the exact true correlation length in the Ising chain in which both the temperature and the field dependencies have been fully expressed. 
	
	\item We have derived the exact scaling functions of the Helmholtz force - see 	\eq{eq:XH-scaling} for periodic, and \eq{eq:antiscaling2}
	for antiperiodic boundary conditions. The behavior of these forces as a function of the scaling variable $x_t$ and magnetization $m$ is shown in Fig. \ref{fig:Helmholtz3D}. The behavior of the force $N \times \beta F_{\rm H}^{(\rm anti)}(N,K,m)\equiv X_{\rm H}^{(\rm per)}(K,m|N)$ under ABC is shown in Fig. 	\ref{fig:Helmholtz} for different values of $N$ and $m$. How this force approaches its scaling behavior given by $X_{\rm H}^{(\rm per)}(K,m)$ is depicted in Fig. \ref{fig:Helmholtz3}. As is clear from these figures - the Helmholtz force changes  sign as a function of the temperature and magnetization and can be attractive or repulsive, depending on their values. 
	
	\item We have shown that for $N\gg 1$ the CE approaches, as expected,  the results which follows from the bulk GCE via Legendre transformation, and we have established that the leading correction to this thermodynamic result in the case of $N$ large, but finite, is of the order of $\ln N/N$ --- see \eq{eq:asymptote}.
	
	\item We have established that the leading corrections to the scaling behavior of the Helmholtz force for $N\gg 1$ are given by a series in terms of $N^{-1}$ --- see \eq{eq:Hfe_scaling} and 	\eq{eq:helmholtz-aper-bc}, as well as  \ref{eq:sec:general_limiting_form}. 
	
	\item In order to achieve our goals we have derived, as a byproduct, several new specific integral representations of the Gauss hypergeometric function, which are of interest on their own. 
\end{itemize}

Finally, we note that the the issue of ensemble dependence of the fluctuation induced force pertinent to a  given ensemble is by no means limited to the Ising chain, or to the canonical ensemble. Calculations, similar to those what we have presented can, in principle, be performed for any ensemble and for any statistical mechanical model of interest. This opens also a very broad front of research for the members of the community performing numerical simulations, say, of the Monte Carlo type.

\section*{Acknowledgements} DD acknowledges the financial support by Grant No BG05M2OP001-1.002-0011-C02 financed by OP SESG 2014-2020 and by the EU through the ESIFs.
NST acknowledges the financial support by 
Grant No D01-229/27.10.2021 of the Ministry of Education and Science of Bulgaria.

\appendix

\section{Derivation of the Henkel's et al. results}
\label{sec:Henkel}

\subsection{Derivation for the case  M=0}
\label{sec:M=0 case}

The partition function for the Ising model defined  on a periodic chain of $\mathcal{N} = 2N$ sites, and magnetization $M=0$ in terms o the Legendre polynomials has the form \cite[p.305, Eq. S4]{Henkel2008}  
\begin{equation}
	\tilde{Z}(K)=(2\sinh(2K))^N[P_N(\coth 2K)-P_{N-1}(\coth 2K)].
\end{equation}
Here we derive this result starting from \eq{eq:Z-final-introduction-per}.

The relation between hypergeometric function and Legendre polynomials  is \cite[p. 456, Eq. 171]{Prudnikov1986}
\begin{equation}
	\label{LG}
	_2F_1(-N,-N;1;z)=(1-z)^{N}\\P_N\left(\frac{1+z}{1-z}\right).
\end{equation}
Differentiating the both sides of Eq.\eqref{LG} we have
\begin{equation}
	\label{derG}
	\frac{d}{dz}\left[_2F_1(-N,-N;1;z)\right]
	=N^2\\_2F_1(1-N,1-N;2;z),
\end{equation}
see \cite[Eq. 15.5.1]{NIST2010}, and, according to  \cite[Eqs. 8.832.1 and 8.832.2]{GR}
\begin{equation}
	\label{lhs}
	\frac{d}{dz} \left[(1-z)^N P_N\left(\frac{1+z}{1-z}\right)\right]=
	\frac{N(1-z)^N}{2z}\left[P_N\left(\frac{1+z}{1-z}\right)-P_{N-1}\left(\frac{1+z}{1-z}\right)\right].
\end{equation}  
If we set $z=e^{-4K} $, from Eqs.\eqref{derG} and \eqref{lhs} 
we get 
\begin{equation}
	\tilde{Z}(K)=2N e^{(2NK-4K)}\; _2F_1(1-N,1-N;2;e^{-4K})=(2\sinh(2K))^N[P_N(\coth 2K)-P_{N-1}(\coth 2K)].
\end{equation}
Remembering that in Ref. \cite{Henkel2008}   the number of sites is an even integer number $2N$,  the above equation establishes the coincidence of the results obtained in both approaches in the case $M=0$.

\subsection{Derivation of our result for the partition function from the Henkel's general result for arbitrary $M$  }
\label{sec:Henkel-general-M}

Here we consider the relation of our result  for the partition function and the Henkel's et al. general result for arbitrary $M$.
For general $M$ and even $N$,  the partition function for the Ising model defined  on a periodic chain of $\mathcal{N} = 2N$ sites has the form \cite[p.307, Eq. S5]{Henkel2008} 
\begin{equation}
	\label{eq:Henkel}
	Z_C^{\rm (per)}(N,K,M)=2Ne^{2KN}\sum_{l=0}^{N}\frac{1}{2N-l}\left( \begin{array}{c} 2N-l \\ N+M/2\end{array} \right) \left( \begin{array}{c} N+M/2 \\ l \end{array} \right)\left(\frac{1-x}{x}\right)^l
\end{equation}
where $x=e^{4K}$. Below we demonstrate how, starting from this expression, one can simplify it to  	\eq{eq:Z-final-introduction-per}  obtained in Ref. \cite{DR2022}.

In terms of $\Gamma(x)$ functions the 	\eq{eq:Henkel} can be rewritten in the form 
\begin{equation}
	Z_C^{\rm(per)}(N,K,M)=2Ne^{2KN}\sum _{l=0}^N t_s(N,M,l) (1-\exp (-4 K))^l,
\end{equation}
where
\begin{equation}
	\label{eq:ts}
	t_s(N,M,l)=\frac{(-1)^l \Gamma (-l+2 N)}{\Gamma (1+l) \Gamma \left(1-l-\frac{M}{2}+N\right) \Gamma
		\left(1-l+\frac{M}{2}+N\right)}. 
\end{equation}
The  ratio in the series that depends on Gamma functions can be simplified to 
\begin{eqnarray}
	\frac{t_s(N,M,l+1)}{t_s(N,M,l)}=
	\frac{\left(l-\frac{M}{2}-N\right) \left(l+\frac{M}{2}-N\right)}{(1+l) (1+l-2 N)}. 
\end{eqnarray}
The first term $ (l=0)$ of the series is
\begin{equation}
	2 N e^{2KN}\frac{\Gamma(2N)}{\Gamma(N+1-M/2)\Gamma(N+1+M/2)}.
\end{equation}
The lookup algorithm then leads to 
\begin{equation}
	\label{H1}
	Z_C^{\rm (per)}(N,K,M)=e^{2 K N} \frac{ \Gamma (1+2 N) }{\Gamma
		\left(1-\frac{M}{2}+N\right) \Gamma \left(1+\frac{M}{2}+N\right)}\, _2F_1\left(\frac{M}{2}-N,-\frac{M}{2}-N;1-2 N;1-e^{-4 K}\right). 
\end{equation} 
Via the Euler hypergeometric transformation formula \cite[9.131.1]{GR} 	\eq{H1} can be rewritten as 
\begin{equation}
	\label{H2}
	Z_C^{\rm (per)}(N,K,M)=\frac{e^{2 K (N-2)} \Gamma (1+2 N) \, }{\Gamma
		\left(1-\frac{M}{2}+N\right) \Gamma \left(1+\frac{M}{2}+N\right)}_2F_1\left(1+\frac{M}{2}-N,1-\frac{M}{2}-N;1-2 N;1-e^{-4 K}\right). 
\end{equation} 
With the help of the indentity \cite[9.131.2]{GR}, from above equation we obtain 	\eq{eq:Z-final-introduction-per} obtained in Ref. \cite{DR2022}.

\section{Derivation of the leading size behavior of the Helmholtz free energy}
\label{appendix:asymptotic}

In the current Appendix we derive the asymptotic behavior of $\beta a^{(\rm per)}(N,K,m)$, see \eq{eq:H-free-energy},
for $N\gg 1$. 



Using the identity
\begin{eqnarray}\label{eq:identity-cosines}
	\cos (\alpha) \cos (\beta) =\frac{1}{2} \left[\cos (\alpha -\beta )+\cos (\alpha +\beta )\right],
\end{eqnarray}
and the fact that $\cos(N m x)$ and $\varphi(x,K)$ are even functions of $x$, from 	\eq{Tch1} we obtain consequently
\begin{eqnarray}\label{eq:Aint}
	{\cal I}(N,K,m) &=& \frac{1}{\pi} \; \int _{-\pi/2}^{\pi/2}\left\{\cos\left[N (m x -\varphi (x,K))\right]+\cos\left[N (m x +\varphi (x,K))\right]\right\}\, \di x\nonumber\\
	&=&\frac{2}{\pi} \; \int _{-\pi/2}^{\pi/2}\cos\left[N (m x -\varphi (x,K))\right]\, \di x = {\rm Re}\left\{\frac{2}{\pi} \; \int _{-\pi/2}^{\pi/2}\exp\left[i N \psi(x|m,K) \right]\, \di x\right\},
\end{eqnarray}
where 
\begin{equation}
	\label{eq:psi-def}
	\psi(x|m,K)\equiv m\, x - \varphi(x,K). 
\end{equation}
For $N$ large, the above integral can be estimated from the saddle-point
	method  \cite[ Eq.(1.7) p.164]{F77}, \cite[Eq. (1.13), p. 264]{Fedoryuk1987}, which in our case amounts to the identity:
	\begin{equation}
		\label{SDM}
		{\cal I}(N,K,m)=\frac{2}{\pi}{\rm Re}\sqrt{-\frac{2\pi}{N\psi''((x_0))}}\exp{[\psi(x_0|m,K)]}\,\left[1+O(N^{-1})\right]
\end{equation}
Next we determine the position  $x_0$ of the saddle-point  of the  function $\psi(x|m,K)$.
We arrive at
\begin{equation}\label{eq:x0}
	x_0=\pm i \sinh ^{-1}\left(\frac{e^{-2 K} m}{\sqrt{1-m^2}}\right)=\pm \, i h_b(K,m),
\end{equation}
where $h_b(K,m)$ is defined in \eq{eq:hb-versus-m}. 
Further we get
	\begin{equation}
		\label{200}
		\psi(x_0|m,K)=i \left[\pm m h_b -\cosh ^{-1}\left(\frac{\cosh (h_b)}{\sqrt{1-e^{-4 K}}}\right)\right]
	\end{equation}
	and
	\begin{equation}
		\label{100}
		\frac{2}{\pi}{\rm Re}\sqrt{-\frac{2\pi}{N\psi''((x_0))}}=\frac{4  }{\sqrt{2 \pi  N \chi_b}} 
\end{equation}
%
%
where we have used that 
\begin{equation}\label{eq:varphi-at-x0}
	\varphi(x_0,K)=	\cos ^{-1}\left(\frac{\cosh (h_b)}{\sqrt{1-e^{-4 K}}}\right)=i \cosh ^{-1}\left(\frac{\cosh (h_b)}{\sqrt{1-e^{-4 K}}}\right),
\end{equation}
and that 
\begin{equation}\label{chi-b-appendix}
	\chi_b\equiv \chi_b(K,m)=e^{2 K} \left(1-m^2\right) \sqrt{1-\left(1-e^{-4 K}\right) m^2}.
\end{equation}
Obviously, in order to be able to apply the saddle  point method we out to use the root $x_0=i h_b(K,m)$. Setting \eqref{200} and\eqref{100} in \eqref{SDM} we get 
\begin{equation}\label{eq:integral-final-form-result}
	{\cal I}(N,K,m)\simeq\frac{4  }{\sqrt{2 \pi  N \chi_b}} \; \exp  \left\{N\left[-  h_b m+  \cosh ^{-1}\left(\frac{\cosh (h_b)}{\sqrt{1-e^{-4 K}}}\right) \right]\right\}\left[1+O(N^{-1})\right].
\end{equation}
Taking into account that
\begin{equation}\label{arccos-representation}
	\cosh ^{-1}(x)=\ln \left(\sqrt{x^2-1}+x\right),
\end{equation}
and, see \eq{eq:hb-versus-m}, 
\begin{equation}\label{eq:helpful}
	\cosh (h_b)=\sqrt{\frac{e^{-4 K} m^2}{1-m^2}+1},
\end{equation}
we obtain 
\begin{equation}\label{eq:arccoh-cos}
	\cosh ^{-1}\left(\frac{\cosh (h_b)}{\sqrt{1-e^{-4 K}}}\right)=\ln \left(\frac{\sqrt{1-\left(1-e^{-4 K}\right) m^2}+e^{-2 K}}{\sqrt{\left(1-e^{-4 K}\right) \left(1-m^2\right)}}\right)
\end{equation}
and, therefore we end up with the asymptotic expansion Eq.\eqref{eq:asymptote} in the main text: 
\begin{equation}\label{eq:integral-final-form-result-in-m}
	{\cal I}(N,K,m) \simeq\frac{4  }{\sqrt{2 \pi  N \chi_b}} \; \exp  \left\{N\left[-  h_b m+  \ln \left(\frac{\sqrt{1-\left(1-e^{-4 K}\right) m^2}+e^{-2 K}}{\sqrt{\left(1-e^{-4 K}\right) \left(1-m^2\right)}}\right) \right]\right\}\left[1+O(N^{-1})\right].
\end{equation}
From here and 	\eq{eq:H-free-energy}for the bulk Helmholtz free energy it is easy to obtain
\begin{equation}\label{eq:the-bulk-results}
	\beta a_b(K,m_b)= m_b\, h_b(K,m_b)-\ln \frac{\sqrt{1-\left(1-e^{-4 K}\right) m_b^2}+e^{-2 K}}{\sqrt{\left(1-e^{-4 K}\right) \left(1-m_b^2\right)}}-\ln [r(K)].
\end{equation}
The last expression can be also written in the form given in \eq{eq:Helmholtz_free_energy_bul_explicit} in the main text. From there, and from \eq{eq:H-free-energy} and  \eq{eq:integral-final-form-result-in-m}, it follows  \eq{eq:corrections-Helmholtz} again given  in the main text. 

\section{	Algebra of  the hypergeometric function of Gauss and the integral representations of the PBC and ABC canonical statistical sums}
\label{section:integral-hypergeometric}

In this Appendix we will obtain the hypergeometric series representations of canonical statistical sums  starting from its integral representations obtained  by transfer matrix approach.  We note at this point the following  important  facts, which explain why we undertake  our study: 

-It is the transfer matrix method which is traditionally preferable  in the literature about one-dimensional Ising thematics. It can be propagated to the case of canonical ensemble as well. 

- The Gauss hypergeometric series: 
\begin{equation}
\label{dhf}
_2F_1\left(\alpha,\beta;\gamma;z\right)=
\sum_{k=0}^{\infty}\frac{(\alpha)_k(\beta)_k}
{(\gamma)_k}\frac{z^k}{k!}, 
\end{equation}
with parameters  $\alpha, \beta, \gamma $, and  $\gamma \neq  0, -1, -2, ... ,$, which in the text  are associated  with $N$ and $M$,
has permeated throughout theory. 
The relations between $\,_2F_1(\alpha,\beta,;\gamma;z) $ and any two  hypergeometric functions with the same argument "z" and with parameters  $\alpha, \beta $ and $\gamma$  changed by $\pm 1$ (named {\it contiguous} functions), can be used  to make a correspondence between statistical sums of Ising chains of different numbers of spins and different boundary conditions.

At the beginning, we will formulate two mathematical identities  which seems are interesting  in their own right.

The following identities are true:

{\bf Identity concerning \eq{eq:Z-final-introduction-per}}

\begin{equation}
\label{eq:what-we-need-to-demonstrate}
{\cal I}(N,K,M)	=c(N,K)\,_2F_1\left(\frac{1}{2}(-M-N+2),\frac{1}{2}(M-N+2);2;e^{-4K}\right),
\end{equation}
where
\begin{equation}
\label{eq:main-integral}
c(N,K)=Ne^{-4K}(1-e^{-4K})^{-N/2},  \qquad {\cal I}(N,K, M):=\frac{4}{\pi}\int_0^{\pi/2}\cos(Mx)\cos [N\varphi(K,x)]\;
dx,
\end{equation}
and $\varphi(K,x)$ is defined by \eqref{eq:varphi-function}.

This statement   follows from Eq.\eqref{frt}, proven below,  since  $I(\alpha,\beta;z) \to I(N,M,K)$ after the substitutions  $\alpha=\frac{1}{2}(-M-N+2)$,  $\beta=\frac{1}{2}(M-N+2)$ and $z=e^{-4K}$.

{\bf Identity concerning \eq{eq:Z-final-introduction-antiper}}
As in the case of PBC we will establish the relation between the expression for the statistical sum Eq.\eqref{eq:Z-final-introduction-antiper} obtained in term of Gauss hyperergeometric functions and the corresponding integral representation Eq.\eqref{ICh2}.
\begin{eqnarray}
\label{eq:introduction-antiper}
&&	{\rm  D}(N,K;M)=\frac{4}{\pi}\int_{0}^{\pi/2}\cos (M \phi) \ x(K,i\phi)U_{N-1}\left[x(K,i\phi)\right]\,d\phi,\nonumber\\
& &= e^{K (N-4)}\left(\sqrt{2\sinh(2K)}
\right)^{-N} \left[2 \left(e^{4 K}-1\right) \, _2F_1\left(\frac{1}{2} (-M-N+2),\frac{1}{2}
(M-N+2);1;e^{-4 K}\right) \right.  \nonumber\\
&&\left.+N \, _2F_1\left(\frac{1}{2} (-M-N+2),\frac{1}{2}
(M-N+2);2;e^{-4 K}\right)\right]. 
\end{eqnarray}

Before proceeding with   the proof of the  identities given by  Eqs. \eqref{eq:what-we-need-to-demonstrate} and \eqref{eq:introduction-antiper}, we introduce  the  more convenient shorthand notations:

{\bf Two basic functions : definitions }

\begin{eqnarray}
\label{aaa}
{\cal I}(\alpha,\beta;z):=\frac{4}{\pi}\int_0^{\pi/2}
\cos\left[(\alpha-\beta)x\right]\cos \left[(2-\alpha-\beta)\;\varphi(z,x)\right]dx
\end{eqnarray}
and
\begin{equation}
\label{bbb}
{\rm  D}(\alpha,\beta;z):=\frac{4}{\pi}\int_0^{\pi/2}\frac{\cos[(\alpha-\beta)x]\cos(x)}{\sqrt{\sin^2(x)-z}}
\sin\left[(2-\alpha-\beta)\;\varphi(z,x)\right]\,dx,
\end{equation}
where 
\begin{equation}
\label{Ob}
\alpha:=\frac{1}{2}(-M-N+2),\quad  \beta:=\frac{1}{2}(M-N+2)\quad  z:=e^{-4K}.
\end{equation}
$\alpha$ and $\beta$  are \textit{nonpositive} \textit{integer} numbers, $\alpha<\beta$, and
\begin{equation}
\label{eq:phi-definition}
\varphi(z,x)=\arccos\left(\frac{\cos(x)}{\sqrt{1-z}}\right).
\end{equation}

We start by formulating two {\it  basic identities A and B}.

{ \bf i) Identity A}

Under the definition:

\begin{equation}
\label{ppp}
{\cal I}(\alpha,\beta,z) :=\frac{4}{\pi}\int_0^{\pi/2}\cos[(\alpha-\beta)x]
T_{2-\alpha-\beta}\left[\frac{\cos(x)}{\sqrt{1-z}}\right]dx,
\end{equation}
where $T_n(x)$ is the Chebyshev polynomials of the first kind,  
the following identity holds:
\begin{eqnarray}
\label{frt}
{\cal I}(\alpha,\beta;z)=(2-\alpha-\beta)\;z\;(1-z)^{-(2-\alpha-\beta)/2}
\,_{2}F_1(\alpha,\beta
;2;z).
\end{eqnarray}

{\bf Proof:}

The  evaluation of $T_n(x)$ in Eq.\eqref{ppp} as a series  is given by \cite[Eq. (5.34)]{mason2002chebyshev}:
\begin{equation}
\label{Tas}
T_n(x)=\frac{n}{2}\sum_{k=0}^{[n/2]}(-1)^k
\frac{(n-k-1)!}{k!(n-2k)!}(2x)^{(n-2k)}, \quad n>0.
\end{equation}
Inserting  Eq.\eqref{Tas} in Eq.\eqref{ppp},
and interchanging the order of integration  and summation, after using the identity \cite[Eq. 3.631.9]{GR} 
\begin{equation}
\label{GR}
\int_0^{\pi/2}\cos(bx)\cos^{\nu-1}(x)dx=\frac{\pi}{\nu 2^\nu B(\frac{1}{2}
	(\nu + b + 1,\frac{1}{2}(\nu -b + 1)},
\quad \mathcal{R}e \; \nu > 0,
\end{equation}
with $b=\beta -\alpha$ and  $\nu=3-\alpha-\beta-2k$,
$B(x,y)=\Gamma(x)\Gamma(y)/\Gamma(x+y)$,
we obtain:
\begin{equation}
\label{Tser3}
{\cal I}(\alpha,\beta;z)=-(2-\alpha-\beta)
\sum_{k=0}^{2-\beta}(-1)^k
\frac{\Gamma(2-\alpha-\beta-k)}
{\Gamma(k+1)
	\Gamma(2-k-\beta)\Gamma(2-k-\alpha)}
(1-z)^
{-(2-\alpha-\beta-2k)/2}.
\end{equation}

The next step is to identify Eq. \eqref{Tser3} as a hypergeometric series (for the hypergeometric series lookup algorithm see, e.g. Ref. \cite{Petkovsek1961})).
It is possible if ratio of the consecutive terms is a rational function of the summation index $k$, which is exactly our case: 
\begin{equation}
\frac{t_{k+1}(z)}{t_{k}(z)}=(-1)\frac{(k+\beta-1)
	(k+\alpha-1)}{(k+1)(k+\alpha+\beta-1)}(1-z).
\end{equation}
Therefor \eq{Tser3} is a Gauss hypergeometric series.
Furthermore one must take into account  that the standardized hypergeometric series begins with a term equal to 1.
Thus from \eq{Tser3}, normalizing by the first term, in virtue of the hypergeometric series lookup algorithm  we obtain for the integral in \eq{ppp} the  result:
\begin{eqnarray}
\label{iI}
{\cal I}(\alpha,\beta;z)={\tilde c}(z)
_2F_1(
\alpha-1,\beta-1;\alpha+\beta-1;1-z), 
\end{eqnarray}
where
$${\tilde c}(z)\equiv (2-\alpha-\beta)\frac{\Gamma(2-\alpha-\beta)}
{\Gamma(2-\beta)\Gamma(2-\alpha)}(1-z)^{-
	(2-\alpha-\beta)/2}.$$

Further, using in Eq.\eqref{iI} the Euler’s transformation formula \cite[Eq. 9.131.1]{GR} in the form
\begin{equation}
	\label{eq:an-identity}
_2F_1(\alpha-1,\beta-1;\alpha + \beta -1;
1-z)=
-z\; _{2}F_1(\alpha,\beta;\alpha+\beta-1;1-z)
\end{equation}
we obtain:
\begin{eqnarray}
\label{iIm}
{\cal I}(\alpha,\beta;z)=c(z)
\,_{2}F_1(\alpha,\beta
;\alpha+\beta-1;1-z),
\end{eqnarray}
where
\begin{equation}
c(z):=(2-\alpha-\beta)\frac{\Gamma(2-\alpha-\beta)}
{\Gamma(2-\beta)\Gamma(2-\alpha)}(1-z)^{-(2-\alpha-\beta)/2}z.
\end{equation}

In order to transform the argument of the hypergeometric function in the rhs of \eq{iIm}, to be $z$ instead of $1-z$  we will use the identity \cite[Eq. 15.87]{NIST2010} 
\begin{eqnarray}
\label{1GRi4}
_2F_1(\alpha,\beta;\alpha + \beta-1;1-z)\,
&=&\,\frac{\Gamma(2-\alpha)\,
	\Gamma(2-\beta)}{\Gamma(2)\Gamma(2-\alpha-\beta)}\, _2F_1(a,b;2;z)
\end{eqnarray}
which is a rearranged version of the formula \cite[Eq. 9.131.2.]{GR}, in which the term with the poles of the Gamma functions in the denominator is canceled.

Thus, after simple calculations we obtain
the final result:
\begin{eqnarray}
\label{iI2N}
{\cal I}(\alpha,\beta;z)=C(z)
\,_{2}F_1(\alpha,\beta
;2;z),
\end{eqnarray}
where
\begin{equation}
\label{iI2Na}
C(z):=(2-\alpha-\beta)(1-z)^{-(2-\alpha-\beta)/2}
z.
\end{equation}
This result complete  the proof.  $\qquad\blacksquare$

{\bf ii) Identity B}

Under the definition
\begin{eqnarray}
\label{babx}
{\rm D}\left(\alpha,\beta;z\right) =
\frac{4}{\pi}\int_{0}^{\pi/2}\cos[(\alpha-\beta) \phi ] \frac{\cos(\phi)}{\sqrt{1-z}}U_{1-\alpha-\beta}\left(\frac{\cos(\phi)}{\sqrt{1-z}}\right)\,d\phi,
\end{eqnarray}
the following identity holds:
\begin{eqnarray}	
{\rm D}(\alpha,\beta,z)=
(1-z)^{-(2-\alpha-\beta)/2}
z\;
\bigg\{2(1/z-1)\, _2F_1(a,b;1;z)
+
(2-\alpha-\beta)\,_2F_1(a,b;2;z)\bigg\}.
\end{eqnarray}

{\bf Proof:}

The  evaluation of $U_n(x)$  as a series  is given by \cite[Eq. (5.34)]{mason2002chebyshev}:
\begin{equation}
\label{Ts}
U_n(x)=\sum_{k=0}^{[n/2]}(-1)^k
\frac{(n-k)!}{k!(n-2k)!}(2x)^{(n-2k)}. \quad n>0
\end{equation}

Inserting  Eq.\eqref{Ts} in \eqref{babx} with $n=1-\alpha-\beta$ we get
\begin{equation}
{\rm D}(\alpha,\beta,z):=\frac{4}{\pi}\int_{0}^{\pi/2}\cos((\beta-\alpha) \phi)  \frac{\cos(\phi)}{\sqrt{1-z}}\sum_{k=0}^{\left[\frac{1-\alpha-\beta}{2}\right]}(-1)^k
\frac{(1-\alpha-\beta-k)!}{k!(1-\alpha-\beta-2k)!}\bigg(2\frac{\cos(\phi)}{\sqrt{1-z}}\bigg)^{(1-\alpha-\beta-2k)}\,d\phi.
\end{equation}
Here, after interchanging the order of integration and summation we get
\begin{equation}
\label{1234}
{\rm D}(\alpha,\beta;z)=\frac{2}{\pi}\sum_{k=0}^{\left[\frac{1-\alpha-\beta}{2}]\right]}(-1)^k
\frac{(1-\alpha-\beta-k)!}{k!(1-\alpha-\beta-2k)!}2^{2-\alpha-\beta-2k}\left[\frac{1}{\sqrt{1-z}}\right]^{2-\alpha-\beta-2k}
\int_{0}^{\pi/2}\cos(M \phi) \left[\cos(\phi)\right]^{2-\alpha-\beta-2k}\,d\phi.
\end{equation}
Now, using the identity \eq{GR} 
with $b=\beta-\alpha$ and  $\nu=3-\alpha-\beta-2k$,
we obtain:
\begin{eqnarray}
\label{GR2}
&&\int_0^{\pi/2}\cos((\beta-\alpha)x)\cos^{2-\alpha-\beta-2k}(x)dx =\frac{\pi}{(3-\alpha-\beta-2k ) 2^{3-\alpha-\beta-2k} B(\frac{1}{2}
	(4-2\alpha-2k),\frac{1}{2}(4-2\beta-2k  ))}\nonumber\\
&&=\frac{\pi}{(3-\alpha-\beta-2k ) 2^{3-\alpha-\beta-2k}}\frac{\Gamma (4-\alpha-\beta-2k)}{\Gamma\left(2-\alpha-k\right)
	\Gamma\left(2-\beta-k\right)}.
\end{eqnarray}
Using the notation $w(z)=\left[1-z\right]^{-1/2}$, after inserting\eq{GR2} into \eq{1234}, we get 
\begin{eqnarray}
\label{11D1}
{\rm D}(\alpha,\beta,z)&=&\sum_{k=0}^{2-\beta}(-1)^k
\frac{\Gamma(2-\alpha-\beta-k)}{k!}
\frac{(2-\alpha-\beta-2k)}
{\Gamma\left(2-\alpha-k\right)
	\Gamma\left(2-\beta-k\right)}\;w(z)^{2-\alpha -\beta-2k}.
\end{eqnarray}
One can rewrite the above equation in the form:
\begin{eqnarray}
\label{1D1}
{\rm D}(\alpha,\beta,z)&=& (2-\alpha-\beta)\, w^{2-\alpha-\beta}(K) 
\sum_{k=0}^{2-\beta}(-1)^k
\frac{\Gamma(2-\alpha-\beta-k)}{k!}
\frac{1}{\Gamma\left(2-\alpha-k\right)
	\Gamma\left(2-\beta-k\right)}\;w^{-2k}(z)\nonumber\\
&&	-2\,w^{2-\alpha-\beta}(z) \sum_{k=0}^{1-\beta}(-1)^{k+1}
\frac{\Gamma(1-\alpha-\beta-k)}{k!}
\frac{1}{\Gamma\left(1-\alpha-k)\right)
	\Gamma\left(1-\beta-k)\right)}\;w^{-2(k+1)}(z).
\end{eqnarray}
Here we have taken into account that for $k>(2-\beta)$ in the first sum, and  $k>(1-\beta)$ in the second, all the terms are zero due to the poles in the corresponding Gamma functions in the denominator. 

The next step is to identify Eq. \eqref{1D1} as a sum of two hypergeometric series (for the hypergeometric series lookup algorithm see, e.g. Ref. \cite{Petkovsek1961})).

If 
\begin{equation}
\label{eq:first-sum-terms}
a(k):=	(-1)^k
\frac{\Gamma(2-\alpha-\beta-k)}{k!}
\frac{1}{
	\Gamma\left((2-\alpha-k)\right)\Gamma\left(2-\beta-k\right)}\;w^{-2k}(z),
\end{equation}
and 
\begin{equation}
\label{eq:first-sum-terms}
b(k):=	(-1)^{k+1}\frac{\Gamma(1-\alpha-\beta-k)}{k!}
\frac{1}{\Gamma\left(1-\alpha-k)\right)
	\Gamma\left(1-\beta-k\right)}\;w^{-2(k+1)}(z),
\end{equation}
we obtain
\begin{itemize}
	\item for the first sum
	\begin{equation}
	\label{eq:first-sum-zero-term-and-ratio}
	a(0)=\frac{\Gamma (2-\alpha-\beta)}
	{\Gamma \left(2-\alpha\right)\Gamma \left(2-\beta\right)}  \quad \mbox{with} \quad \frac{a(k+1)}{a(k)}=\frac{\left(k-1+\beta\right) \left(k-1+\alpha\right)}{(k+1)  (k+1-\alpha-\beta)}\; w^{-2}(z),
	\end{equation}
	and 
	\item for the second sum 
	\begin{equation}
	\label{eq:first-sum-zero-term-and-ratio}
	b(0)=-\frac{\Gamma (1-\alpha-\beta)}{\Gamma \left(1-\alpha\right) \Gamma \left(1-\beta\right) }
	\;w^{-2}(z)\quad \mbox{with} \quad \frac{b(k+1)}{b(k)}=\frac{\left(k+\alpha\right) \left(k+\beta \right)}{(k+1) (k-\alpha-\beta)} \; w^{-2}(z).
\end{equation}
Thus, applying the lookup algorithm, we obtain
\item for the first sum 
\begin{equation}
\label{eq:first-sum-result}
(2-\alpha-\beta)\, w(z)^{(2-\alpha-\beta)}\frac{ \Gamma (2-\alpha-\beta) }{\Gamma \left(2-\alpha\right)\Gamma \left(2-\beta\right) }\; _2F_1\left(\alpha-1,\beta-1;\alpha + \beta - 1;w^{-2}(z)\right)
\end{equation}
and 
\item for the second sum 
\begin{equation}
\label{eq:second-sum-result}
2w(z)^{(1-\alpha-\beta)}\frac{ \Gamma (1-\alpha-\beta) }{\Gamma \left(1-\alpha\right)\Gamma \left(1-\beta\right) }\, _2F_1\left(\alpha-1,\beta - 1;\alpha+\beta;w^{-2}(z)\right).
\end{equation}
\end{itemize}
Putting all together, using \eq{eq:an-identity}, we arrive at 
\begin{eqnarray}
\label{D}
{\rm D}(\alpha,\beta,z)&=&
(1-z)^{-(2-\alpha-\beta)/2}\frac{\Gamma(1-\alpha-\beta)}
{\Gamma\left(1-\alpha\right)
\Gamma\left(1-\beta\right)}
\bigg \{ [1-z]\, _2F_1 \left(\alpha ,\beta;\alpha+\beta;1-z\right)\nonumber\\ 
&& +
2\frac{(1-\alpha-\beta)(2-\alpha-\beta) }{(2-\alpha-\beta)^2-(\alpha-\beta)^2} \,    
_2F_1 \left(\alpha ,\beta;\alpha +\beta - 1;1-z\right)
\bigg\}
\end{eqnarray}

In order to transform the argument of the hypergeometric function in the rhs of \eq{iIm}, to be $z$ instead of $1-z$  we will use the identities \cite[Eq. 15.87]{NIST2010} 
\begin{eqnarray}
\label{1GR}
_2F_1(\alpha,\beta;\alpha + \beta-1;1-z)\,
&=&\,\frac{\Gamma(2-\alpha)\,
\Gamma(2-\beta)}{\Gamma(2-\alpha-\beta)}\, _2F_1(a,b;2;z)
\end{eqnarray}
and
\begin{eqnarray}
\label{2GR}
_2F_1(\alpha,\beta;\alpha + \beta;1-z)\,
&=&\,\frac{\Gamma(1-\alpha)\,
\Gamma(1-\beta)}{\Gamma(1-\alpha-\beta)}\, _2F_1(a,b;1;z)
\end{eqnarray}
which are obtained from the formula \cite[Eq. 9.131.2.]{GR}, in which the term with the poles of the Gamma functions in the denominator is canceled.
Plugging Eqs. \eqref{1GR} and \eqref{2GR} into Eq.\eqref{D} we get
\begin{eqnarray}
{\rm D}(\alpha,\beta,z)&=&
(1-z)^{-(2-\alpha-\beta)/2}\frac{\Gamma(1-\alpha-\beta)}
{\Gamma\left(1-\alpha\right)
\Gamma\left(1-\beta\right)}
\bigg \{ [1-z]\, \,\frac{\Gamma(1-\alpha)\,
\Gamma(1-\beta)}{\Gamma(1-\alpha-\beta)}\, _2F_1(a,b;1;z)\nonumber\\ 
&&+
2\frac{(1-\alpha-\beta)(2-\alpha-\beta) }{(2-\alpha-\beta)^2-(\alpha-\beta)^2}    
\,\frac{\Gamma(2-\alpha)\,
\Gamma(2-\beta)}{\Gamma(2-\alpha-\beta)}\, _2F_1(a,b;2;z)
\bigg\}
\end{eqnarray}
or, simplifying
\begin{eqnarray}
\label{123}
{\rm D}(\alpha,\beta,z)=
(1-z)^{-(2-\alpha-\beta)/2}
z\;
\bigg\{2(1/z-1)\, _2F_1(a,b;1;z)
+
(2-\alpha-\beta)\,_2F_1(a,b;2;z)\bigg\},
\end{eqnarray}
which is the  result required.  $\qquad\blacksquare$

{\bf Corollary 1:}

Now we are in a position to establish the validity of \eq{eq:Z-final-introduction-antiper} from Ref. \cite{DR2022}. 
The proof  follows directly from the identity Eq.\eqref{123} after the replacing  $\alpha=\frac{1}{2}(-M-N+2),\quad \beta=\frac{1}{2}(M-N+2),\quad z=e^{-4K}$ and the definition of the statistical sum Eq.~\eqref{eq:transfer-matrix-antiperiodic-final-simple}. 

{\bf Corollary 2:}

With the help of the {\it contiguous} relation (see Eq. (2.25) in Ref. \cite[]{Rakha2011} with  $c=2$)
\begin{equation}
\label{eq:identity-F1-F2p}
\,_2F_1(\alpha,\beta;1;z)=-\frac{\alpha\beta}{1-\alpha-\beta}(1-z)\; _2F_1(\alpha+1,\beta+1;2;z)
+
\frac{(1-\alpha)(1-\beta)}{1-\alpha-\beta}\; _2F_1(\alpha,\beta;2;z),
\end{equation}
after some algebra one can transform Eq.\eqref{123}, and the related result is
\begin{eqnarray}
\label{DDDg}
{\rm D}(\alpha,\beta;z)&=& 
(1-z)^{-(2-\alpha-\beta)/2} \left\{
\left[[2\,-\,(\alpha + \beta)z]-\alpha\beta
\frac{2(1-z)}{(\alpha+\beta -1)}\right]\,_2F_1(\alpha, \beta;2;z) \right.\nonumber\\
&& \left. +
\alpha\beta 
\frac{2(1-e^{-4K})^2}{(\alpha+\beta-1)}\,_2F_1(\alpha+1, \beta+1;2;z)\right\}.
\end{eqnarray}
Using Eq.\eqref{DDDg}, after plugging there $\alpha=\frac{1}{2}(-M-N+2),\quad \beta=\frac{1}{2}(M-N+2)$ and $z=e^{-4K}$, the expressions of $Z^{(\rm per)}_C(N,K,M)$,  \eq{eq:Z-final-introduction-per}, together with that one of $Z^{(\rm anti)}_C(N,K,M)$, see \eq{eq:Z-final-introduction-antiper}, we prove Eq.\eqref{zzantiper} from the main text.

\section{The general limiting form of the partition function in the canonical ensemble in the scaling regime}  \label{eq:sec:general_limiting_form}

The partition function of the one-dimensional Ising model in the canonical ensemble is expressed in terms of hypergeometric functions. In the current Appendix  we determine the behavior of the hypergeometric functions in the scaling regime  ($N \gg 1, e^{2K} \sim N$).

First, we recall that if in \eq{dhf} $\alpha$ and/or $\beta$ are
negative integers, which is the case of the present study, 
the Gauss series reduces to a hypergeometric polynomial.  Explicitly, 
\begin{equation}
	_2F_1(-a,-b;c;z) = \sum_{n=0}^{\infty} \frac{(-a)_n(-b)_n}{(c)_n} \frac{z^n}{n!}  \label{eq:F21-def}, \quad \{a,b,c>0\}, \in \mathbb{N}.
\end{equation}
Using the Pochhammer's symbol
\begin{equation}
	\label{DePo}
	(-p )_n= \left\{\begin{array}{cc}n!(-1)^n\binom p n, & 1\leq n \leq p\\  
		\\0, & \qquad n\geq p+1, \qquad \end{array}\right., \quad p,n \in \mathbb{N},
\end{equation}
one obtains
\begin{equation}
	\label{eq:F21-d}
	_2F_1(-a,-b;c;z) = \sum_{n=0}^{b-1} n!\binom a n  \binom b n\frac{(c-1)!}{(c+n-1)!}z^n , \quad \{a,b,c>0\} \in \mathbb{N}.
\end{equation}
Here we assumed that $b \leq a$. Using the relation
\begin{equation}
	\binom a n n!=\frac{(a+n-1)!}{(a-1)!},\quad n\in \mathbb{N},
\end{equation}
\eq{eq:F21-d} becomes
\begin{equation}
	_2F_1(-a,-b;c;z) = \frac{(c-1)!}{(a-1)!(b-1)!}\sum_{n=0}^{b-1} \frac{(a+n-1)!(b+n-1)!}{(c+n-1)!} \frac{z^n}{n!}  \label{eq:F21-def-next-step}.
\end{equation}
In the scaling region we are interested in the behavior of the hypergeometric function when $|a|\sim N$ and $|b|\sim N$ are very large compared to 1,  $c$ has a moderately small value,  and $0 \leq z \ll 1$.
Thus, we are interested in the behavior of this function when $a, b \gg 1$. Our main statement in this case is
that 
the following asymptotic expansion  is valid:
\begin{equation}
	\label{I}	
	_2F_1(-a,-b;c;z) \simeq (c-1)! \left(\sqrt{abz} \right)^{1-c}
	\left[I_{c-1}\left(2 \sqrt{abz}\right)+\frac{1}{2}(a+b)\,z\;I_{c+1}\left(2 \sqrt{abz}\right)+O(N^{-2})\right] ,
\end{equation}
where $I_{\nu}(z)$ is the modified Bessel function of first kind.  Actually, the leading term in this expansion, propositional to $I_{c-1}$ originates from the Hansen \cite[Eq. 5.71, p.154]{Watson1944}. Below we explain how to derive this improved result (suitable  for finite $a$ and $b$). We start, when $a\gg 1$, with the known expansion \cite[Eq. 5.11.13]{NIST2010}
\begin{equation}
	\label{AFa}
	\frac{ (a+n-1)!}{(a-1)!}\simeq  a^n\left[1+\sum_{k=1}^{\infty}\frac{G_k(n)}{a^k}\right], \quad 
	\mbox{where}
	 \quad
	G_k(n):=\binom{n}{k} B_k^{n-1}(n). 
\end{equation}
Here  $ B_k^{n-1}(n)$
are generalized Bernoulli polynomials. Explicitly, one has \cite[5.11.17]{NIST2010}:
\begin{equation}
	G_0(n)=1,\quad G_{1}(n)=1 + \frac{1}{2}(n-1)n,\quad 
\end{equation}	
Thus for the ratios of factorials in Eq.\eqref{eq:F21-def-next-step}, for $n$ not too large, using \eq{AFa} one gets (up to the second order in $N^{-1}$)
\begin{equation}
	\label{rb}
	\frac{(a+n-1)!}{(a-1)!} \simeq a^n  \left[1+\frac{(n-1) n}{2 a}+ O(a^{-2})\right] ,
\end{equation}
and similarly for the corresponding ratio for the variable $b$.
For $b\gg 1$,
after using the replacement $\sum_{n=0}^{b-1}\quad \to \quad \sum_{n=0}^\infty$, 
we can approximate the right hand side of Eq.\eqref{eq:F21-def} by
\begin{equation}
	\label{Ru}
	(c-1)! \sum_{n=0}^{\infty} \frac{(z)^n}{(c+n-1)! n!}a^nb^n  \left[1+\frac{(n-1) n}{2 a}\right]  \left[1+\frac{(n-1) n}{2 b}\right]. 
\end{equation}
Recalling  the definition of the modified Bessel function of first kind \cite[10.25.2]{NIST2010}
\begin{equation}
	\label{DefB}
	I_{\nu}(x)=\sum_{k=0}^{\infty}
	\frac{1}{k! \Gamma(\nu+k+1)}\left(\frac{x}{2}\right)^{\nu+2k} \simeq\left\{\begin{array}{cc}\left(\frac{x}{2}\right)^{\nu}/\Gamma(\nu+1), & z\to 0\\ 
		&\\e^x/\sqrt{2\pi x}, & x
		\to \infty, \end{array}\right.,
\end{equation}
and by the virtue of  Eq.\eqref{DefB} we obtain the result stated in 	\eq{I}	.

{\bf Corollary 1:} It can be readily verified that in the scaling regime  the approximation in \eq{eq:F21-def} holds for the terms with moderately large values of $n$. This is because at large values of $n$, the behavior of the summand in \eq{Ru} is controlled by the combination $abz$, where in our case $z=e^{-4K} \ll 1$, while $a$ and $b$ are of order $N$. Note that $z$ is of order $N^{-2}$. Then, the replacement of the sum with infinite series is acceptable due to  the terms $z^n/ n! $, which effectively cuts the high-order terms in the sum over $n$.  

{\bf Corollary 2:}
If we choose $c=1$ and replace in \eq{I} $a$ by $(M+N-2)/2$, and $b$ by 
$(-M+N-2)/2$, we arrive at \eq{eq:scaling-forms-second}, with $N>>1$ and  $z=e^{-4K} \ll 1$.
Let us note that for $c=2$ the property stated from \eq{I} can be also alternatively derived as shown in Eqs.(\ref{eq:Z-in-terms-of-Np-Nm}) -- (\ref{eq:Z-in-terms-of-Np-Nm-to-scaling-second}). 


From Eqs. \eqref{eq:Hfe_scaling}  and \eqref{eq:helmholtz-aper-bc}, one observes that the leading corrections in $N$ of the scaling behavior of the Helmholtz free energy densities $a_s^{(\zeta)}(N,K,m)$   are given by $\ln (N)/N$, while the next terms can be obtained according to Eq. \eqref{I}. Further corrections can be obtained using   Eq.\eqref{AFa}. 

%

	
\end{document}